\documentclass[12pt,pdf]{iopart}
\usepackage{iopams}
\usepackage{graphicx}   

\newcommand{\imag}{{\rm \,Im\,}}
\begin{document}

\title[Fast multi-orbital equation of motion impurity solver for DMFT]{Fast multi-orbital equation of motion impurity solver for dynamical mean field theory}

\author{Qingguo Feng and P. M. Oppeneer}
\address{Department of Physics and Astronomy, Uppsala University, Box 516, S-75120 Uppsala, Sweden}
\ead{qingguo.feng@physics.uu.se}

\date{\today}
\begin{abstract}
We propose a fast multi-orbital impurity solver for the dynamical mean field theory (DMFT). Our DMFT solver is based on the equations of motion (EOM) for local Green's functions
and constructed by generalizing from the single-orbital case to the multi-orbital case with inclusion of the inter-orbital hybridizations and applying a mean field approximation to the inter-orbital Coulomb interactions. The two-orbital Hubbard model is studied using this impurity solver within a large range of parameters. The Mott metal-insulator transition and the quasiparticle peak are well described. A comparison of the EOM method with the QMC method is made for the two-orbital Hubbard model and a good agreement is obtained. The developed method hence holds promise
as a fast DMFT impurity solver in studies of strongly correlated systems.

\end{abstract}

\maketitle

\section{Introduction\label{sect1}}

Since the introduction of the dynamical mean field theory (DMFT)~\cite{DMFT1,DMFT2} in the past decade,
it has been continuously developed and improved and has become a powerful tool to investigate strongly correlated systems involving $d$ and $f$ electrons.
DMFT greatly helps to understand, e.g.,  the Mott metal-insulator transition as well as other exotic properties in strongly correlated electron systems.
The basic idea of the DMFT is to map interactions between different sites to interactions between an impurity and a self-consistently determined bath \cite{DMFTRMP96}.
Then, a lattice problem is transformed to solving a single impurity problem along with a DMFT self-consistency condition. Thus, the central part of DMFT calculation is to solve the impurity self-energy efficiently. This is a numerical problem, to which a considerable amount of effort has been devoted to advance and improve it during the last decade (see, e.g., \cite{DMFTRMP06} for a review).

Frequently used impurity solvers are the quantum Monte Carlo (QMC) method \cite{HF-QMC}, the exact diagonalization (ED) method \cite{ED,ED2}, numerical renormalization group (NRG) method \cite{NRG,NRG2}, density matrix renormalization group (DMRG) method \cite{DMRG}, and recent continuous time quantum Monte Carlo (CTQMC) method \cite{CTQMC,CTQMC2}. But all of these methods are either computationally expensive or work only in a limited parameter region. Therefore, novel impurity solvers that are both fast and reliable are demanded, particularly when calculations for real materials are performed, merging density functional theory (DFT) plus DMFT approaches.
The equations of motion (EOM) method and associated methods have been studied a lot in the past half century
 \cite{Hubbard63,Hubbard64a,Hubbard64b,Lacroix1, Lacroix2,Czycholl85,Petru93,GrosEOM,Kang95,Luo99,ZhuEOM1,ZhuEOM2,OPM1,OPM2}, and has been considered to be a good candidate for a fast impurity solver.
In \cite{JK05} and \cite{FZJ09}, an infinite $U$ and a finite $U$ impurity solver were successfully developed for the single-orbital case based on equations of motion of the Green's functions. Another recent work by Zhuang {\it et al.} exploring multi-orbital impurity solvers is based on the Gutzwiller variational approach \cite{Dai}.

The EOM impurity solver has previously been developed for the single orbital case \cite{FZJ09,phdthesis-feng}. However, in calculations of real materials involving strongly correlated electrons, one usually has orbital degrees of freedom, i.e., near the Fermi surface there exists more than one orbital. For example, when studying transition metal oxides and lanthanides, the valence $d$ (and $f$) electrons usually occupy several, up to 5 (and 7) orbitals. These orbitals have similar energies and, consequently, all of these orbitals have to be considered in the manifold of localized states.
Therefore, it is essential to develop the impurity solver from the single-orbital case to a multi-orbital EOM (MO-EOM) impurity solver that is applicable in concrete DFT+DMFT calculations for real materials.

For QMC, ED and CTQMC methods multi-orbital solvers have been developed. However, according to the earlier investigation for the single orbital case \cite{FZJ09,phdthesis-feng}, the EOM impurity solver is computationally much faster compared to QMC and NRG methods, because the EOM method does not need matrix calculations, but solves the integral equations. As in general calculations of multi-orbital systems will be much more time consuming than those of single-orbital systems, due to the increase of the dimension of matrix, the MO-EOM impurity solver will hopefully save more cpu time. Moreover, it has the advantage that it works directly on real frequencies and can hence be employed for an arbitrary finite temperature.

Our aim is to obtain a fast, reliable, and general multi-orbital impurity solver with EOM method.
Comparing to the single-orbital system with SU({\it 2N}) symmetry which can approximately deal with the $N$-orbital systems with degenerated energies, in the multi-orbital impurity solver all the interactions of the localized electrons should take into account the orbital degrees of freedom and all the orbitals may also variate in their physical parameters, e.g., band widths, intra-orbital and inter-orbital Coulomb interaction strengths, and so on. With more flexibility in the interactions, a wider range of physically relevant cases and more complex systems can be treated, and the method can provide more interesting information.

Furthermore, in all previous studies of the multi-orbital Hubbard model, for the hopping term between different sites, the electrons are all treated such that they only hop to the orbitals with identical orbital indices on other sites, thus the inter-site inter-orbital hoppings are neglected. We think that these inter-site inter-orbital hoppings are important because only when all inter-site hoppings combined in one are included, can the model accurately reflect a realistic physical image of multi-orbital systems. Especially when the orbitals have different band widths  the neglect of inter-site inter-orbital hopping introduces a man-made difference.
Therefore, we have tried to incorporate this inter-site inter-orbital hopping effect in our treatment to study the orbital selective Mott transition (OSMT) in a fully self-consistent way.

The paper is organized as follows:  in section~\ref{sect2} we describe our MO-EOM method and introduce the equations, decoupling, and approximations.
In section~\ref{sect3} we present model calculation results for the MO-EOM by applying it to two-orbital Hubbard model and we compare the computed results with those obtained with the QMC method.
Our findings are summarized in section~\ref{sect4}.

\section{Description of MO-EOM method\label{sect2}}
We start from the Hamiltonian of a modified multi-orbital Hubbard model. It is given as
\begin{eqnarray}
{\cal H}&=&-\sum_{ijlm\sigma,i\neq j}t_{ijlm}f^{\dag}_{il\sigma}f_{jm\sigma}
+\sum_{il}U_{ll}\hat{n}_{il\uparrow}\hat{n}_{il\downarrow}+\sum_{ilm\sigma\sigma',l<m}U_{lm\sigma\sigma'}\hat{n}_{il\sigma}\hat{n}_{im\sigma'}\nonumber\\
&&+\sum_{ilm\sigma,l<m}\big(V'^{\ast}_{lm\sigma}f^{\dag}_{im\sigma}f_{il\sigma}^{~}+V'_{lm\sigma}f^{\dag}_{il\sigma}f^{~}_{im\sigma}\big),\label{eq:1}
\end{eqnarray}
where $i,j$ are site indices, $l,m$ are orbital indices, and $\sigma, \sigma'$ are spin indices, 
respectively. The first summation, which sums over two orbital indices and which differs from the usually studied form of the multi-orbital Hubbard model \cite{OSMT1}, is the hopping of electrons between different sites. The second (third) summation term is the on-site intra-orbital (inter-orbital) Coulomb interaction term, where the intra-orbital Coulomb interaction strength $U_{ll}$ and the inter-orbital Coulomb strength $U_{lm\sigma\sigma'}$ are free parameters that can be set all identical or all different in order to simulate different kinds of physical systems. The last one is the on-site hopping of electrons between different orbitals which should
imperatively be taken into account when multi-orbital effects are studied. Here $V'^{\ast}_{lm\sigma}$ and $V'_{lm\sigma}$ are the on-site inter-orbital hopping parameters.

In the dynamical mean field theory, the many body interactions between different sites are mapped to interactions between
an impurity and a bath. The most frequently used impurity model is the single impurity Anderson model (SIAM).
The Hamiltonian of the multi-orbital SIAM is
\begin{eqnarray}
{\cal H}_{imp}&=&\sum_{kl\sigma}\varepsilon_{kl\sigma}c^{\dag}_{l k\sigma}c_{l k\sigma}^{~}+\sum_{l\sigma}\varepsilon_{fl\sigma} f^{\dag}_{l\sigma}f_{l\sigma}+\sum_{l}U_{ll}\hat{n}_{l\uparrow}\hat{n}_{l\downarrow}\nonumber\\
&+&\sum_{lm\sigma\sigma',l<m}U_{lm\sigma\sigma'}\hat{n}_{l\sigma}\hat{n}_{m\sigma'}\nonumber\\
&+&~~\sum_{l k\sigma}~~~~\big(V^{\ast}_{l k\sigma}c^{\dag}_{l k\sigma}f_{l\sigma}+V_{l k\sigma}f^{\dag}_{l\sigma}c_{l k\sigma}\big)\nonumber\\
&+&\sum_{lm\sigma,l<m}~\big(V'^{\ast}_{lm\sigma}f^{\dag}_{m\sigma}f_{l\sigma}+V'_{lm\sigma}f^{\dag}_{l\sigma}f_{m\sigma}\big).
\label{eq:2}
\end{eqnarray}
Here $\hat{n}_{l\sigma}=f^{\dag}_{l\sigma}f_{l\sigma}^{~}$ is the occupation number for localized electrons with spin $\sigma$ in the $l$-th orbital.
The first summation term is the energy of conduction electrons, where the electrons in different
orbitals are labeled with orbital index $l$.
The second summation term is the energy of localized electrons and $\varepsilon_{fl\sigma}$ is the orbital level for spin $\sigma$ in the $l$-th orbital. The third summation term is the on-site intra-orbital Coulomb interaction term. The fourth summation term is the on-site inter-orbital Coulomb interactions between electrons of the $l$-th orbital and $m$-th orbital. The fifth summation term is the hybridization between the localized electrons and the baths. The sixth summation has the same meaning as in \eref{eq:1}.

The temperature dependent retarded two-time Green's function in the Zubarev
notation~\cite{Zubarev} is given by
\begin{eqnarray}
G_{AB}(t,t')
&=&\ll A(t);B(t')\gg\nonumber\\
&=&-i\Theta(t-t')\langle
[A(t),B(t')]_+\rangle,
\label{eq:3}
\end{eqnarray}
where $A(t)$ and $B(t')$ are Heisenberg operators,
and $\Theta(t-t')$ is the Heavyside function.
Here we use the Fourier transform of the Green's function in $\omega$ space,
it  satisfies the equations of motion
\begin{eqnarray}
&&\omega\ll A;B\gg=\langle[A,B]_+\rangle+\ll[A,{\cal H}_{imp}];B\gg,\label{eq:4}\\
&&\omega\ll A;B\gg=\langle[A,B]_+\rangle+\ll A;[{\cal H}_{imp},B]\gg.\label{eq:4b}
\end{eqnarray}
In the following calculations of the equations of motion \eref{eq:4} is applied.

For a multi-orbital system, if we define the anti-commutation relation for the operators as follows,
\begin{eqnarray}
&&[f_{l\sigma}, f_{m\sigma'}]_+=0, ~[f^{\dag}_{l\sigma},
f^{\dag}_{m\sigma'}]_+=0, \nonumber\\
&&[f^{\dag}_{l\sigma}, f_{m\sigma'}]_+=\delta_{lm}\delta_{\sigma\sigma'},
~[f_{l\sigma}, f^{\dag}_{m\sigma'}]_+=\delta_{lm}\delta_{\sigma\sigma'},\nonumber
\end{eqnarray}
we can obtain the first two equations of motion for single particle Green's functions,

\begin{eqnarray}
\fl (\omega+\mu-\varepsilon_{fm\sigma})\ll
f_{m\sigma};f^{\dag}_{m\sigma}\gg
&=&1+U_{mm}\ll\hat{n}_{m\sigma'}f_{m\sigma};f^{\dag}_{m\sigma}\gg\nonumber\\
&+&\sum_{l,l\neq m}\big(U_{lm\sigma\sigma}\ll\hat{n}_{l\sigma}f_{m\sigma};f^{\dag}_{m\sigma}\gg\nonumber\\
&&\qquad+U_{lm\sigma'\sigma}\ll\hat{n}_{l\sigma'}f_{m\sigma};f^{\dag}_{m\sigma}\gg\big)\nonumber\\
&+&\sum_{k}~\big(V_{mk\sigma}\ll c_{mk\sigma};f^{\dag}_{m\sigma}\gg\nonumber\\
&&\qquad-\sum_{l,l\neq m}V'_{lm\sigma}\ll f_{l\sigma};f^{\dag}_{m\sigma}\gg\big) ,
\label{eq:6}\\
\fl (\omega+\mu-\varepsilon_{km\sigma})\ll
c_{mk\sigma};f^{\dag}_{m\sigma}\gg&=&
{\cal V}^{\ast}_{mmk\sigma}\ll
f_{m\sigma};f^{\dag}_{m\sigma}\gg+\sum_{l,l\neq m}{\cal V}^{\ast}_{mlk\sigma}\ll f_{l\sigma};f^{\dag}_{m\sigma}\gg.
\label{eq:7}
\end{eqnarray}

Here $\mu$ is the chemical potential, $\sigma$ and $\sigma'$ are spin indices where by notation we imply $\sigma'\neq\sigma$. This notation is employed in the following derivation, too.
On the right hand side (RHS) of \eref{eq:6}, the physical meaning of the second term reflects the fluctuation of spin $\sigma$ in $m$-th orbital accompanied by spin $\sigma'$ in $m$-th orbital, or we can call it the fluctuation of spin $\sigma$ when spin $\sigma'$ exists. The third term shows the fluctuation of spin $\sigma$ in $m$-th orbital when spin $\sigma$ in $l$-th orbital exists, and similarly the fourth term is for the case when $\sigma'$ in $l$-th orbital exists. The first term
(i.e., 1) reflects the existence of spin $\sigma$ in the $m$-th orbital itself. To describe the multi-orbital effect, we have to consider each term for the case that other spin-orbital channels exist at the same time. Moreover, \eref{eq:7} describes the hopping of electrons from the bath to the impurity here. The electrons can hop back to any orbital of the impurity. Therefore, from Eqs.~\eref{eq:6} and \eref{eq:7}, we can imagine that the electrons in the bath which came from the $m$-th orbital of the impurity hop to all orbitals of the impurity, where the ${\cal V}^{\ast}_{mmk\sigma}$ and ${\cal V}^{\ast}_{lmk\sigma}$ are the hybridizations of electrons in the bath which came from $m$-th orbital to $m$-th and $l$-th orbital accordingly. There exists the relation that $V^{\ast}_{mk\sigma}={\cal V}^{\ast}_{mmk\sigma}+\sum_{l,l\neq m}{\cal V}^{\ast}_{lmk\sigma}$. This constitutes indirect hybridizations between different orbitals through the bath, which is generated from the bath-impurity hybridization terms of \eref{eq:2} and hence, it reflects the inter-site hopping matrix $t_{ijlm}$ in the Hamiltonian of the Hubbard model, \eref{eq:1}. This is the reason that we introduce the inter-site hopping matrix for multi-orbitals in the Hubbard model. But one can recognize that ${\cal V}^{\ast}_{mmk\sigma}$ and ${\cal V}^{\ast}_{lmk\sigma}$ will not appear in the impurity Hamiltonian \eref{eq:2} because for indirect inter-orbital hybridizations each orbital of the impurity can only "see" the bath. These diagonal and off-diagonal hybridization parameters can only be obtained from the mapping of the inter-site hopping on the lattice model.

In the EOM method when the equations of motion are derived, a decoupling scheme is implemented to make the equations closed and solvable.
However, this decoupling scheme is not only a simplifying numerical technique. We emphasize here that it is based on different grades of realistic physical assumptions. In the EOM method the interactions included are not simply considered as the interactions shown in the model Hamiltonian, but they also depend on the included equations of motion. In calculating the equations of motion, more and more higher order Green's functions will appear in higher and even higher order equations of motion. These higher Green's functions are associated with some kind of higher order interactions. From the order of the equations of motion in which a Green's function first appears, one can define the order of the Green's functions and approximately evaluate the contribution of the related interactions because this `order' approximately describes the weight of an interaction. For two Green's functions with different order, the lower order Green's function associated with an interaction will have a larger weight and hence give more contribution than the higher order Green's function associated with the interaction. This is approximately valid because the contribution will also depend on the interaction strengths (e.g., the values of $U$, $J$, $V$, and so on),
and not only on the order. Once one higher order Green's function appears, one new higher equation of motion can be derived, while for one new equation of motion more higher order Green's functions will appear. This is an infinite procedure. The decoupling scheme implies to make a decision where to truncate this procedure and only the interactions before this truncation will be exactly taken into account and all the interactions higher than this truncation will be approximated with lower order Green's functions and relative correlation functions. Therefore, the decoupling scheme decides how many interactions are fully included. Understandably, a higher order decoupling scheme will take more interactions exactly into account and will be more accurate. This is the physical meaning of the decoupling scheme, and it is also the reason that we call one interaction a {\it higher order} interaction than another interaction, by just comparing the orders of the two Green's functions associated with the interaction.
Thus inclusion of higher-order interactions is a controllable procedure. Therefore, the EOM method is a method that can infinitely approach the exact physics, whose accuracy depends on the order of the equations of motion or the number of interactions that have been exactly included.

To illustrate this we adopt the single orbital system \cite{FZJ09} as an example to show the physical implications of the decoupling scheme. For the lowest order decoupling which corresponds to the mean field approximation for the Coulomb interactions, all the electrons in one orbital are considered moving in the mean field of other electrons, where only the on-site Coulomb interactions and the single-electron hybridization interaction $f^{\dag}_{\sigma}c_{k\sigma}$ are fully taken into account.
Consequently $\ll\hat{n}_{\sigma}f_{\sigma};f^{\dag}_{\sigma}\gg$ can be decoupled as $\bar{n}_{\sigma}\ll f_{\sigma};f^{\dag}_{\sigma}\gg$ and only two equations of motion need to be derived,
\begin{eqnarray}
(\omega+\mu-\varepsilon_f)\ll f_{\sigma};f^{\dag}_{\sigma}\gg&=&1+U\ll \hat{n}_{\sigma'}f_{\sigma};f^{\dag}_{\sigma}\gg+V_{k\sigma}\ll c_{k\sigma};f^{\dag}_{\sigma}\gg,\nonumber\\
(\omega+\mu-\varepsilon_k)\ll f_{\sigma};f^{\dag}_{\sigma}\gg&=&V^{\ast}_{k\sigma}\ll f_{\sigma};f^{\dag}_{\sigma}\gg,\nonumber
\end{eqnarray}
where $\ll\hat{n}_{\sigma'}f_{\sigma};f^{\dag}_{\sigma}\gg$ and $\ll c_{k\sigma};f^{\dag}_{\sigma}\gg$ are the same order Green's functions because they appear simultaneously, and are corresponding to the two most basic interactions mentioned above.
Therefore, the single particle Green's function should be
\begin{eqnarray}
\ll f_{\sigma};f^{\dag}_{\sigma}\gg=\frac{1}{\omega+\mu-\varepsilon_f-\Delta-U},
\end{eqnarray}
which gives only an energy shift of the band along with the Coulomb interaction.

For the equivalent of the Hubbard-I approximation which is the second order EOM decoupling scheme (here we call this the equivalent Hubbard-I approximation because the Hubbard-I approximation is originally for the Hubbard model but not for SIAM), one more four-operator single-electron hybridization interaction $\hat{n}_{\sigma'}f^{\dag}_{\sigma}c_{k\sigma}$ is exactly taken into account, which relates to the Green's function $\ll\hat{n}_{\sigma'}c_{k\sigma};f^{\dag}_{\sigma}\gg$. This interaction can be viewed as one next higher order interaction than the Coulomb interactions because the Coulomb interaction first appear in the first order equations of motion and this interaction first appears in the second order equations of motion. In this case four equations of motion will be needed, and finally lead to
\begin{eqnarray}
\ll f_{\sigma};f^{\dag}_{\sigma}\gg=\frac{1-\bar{n}_{\sigma'}}{\omega+\mu-\varepsilon_f-\Delta}+\frac{\bar{n}_{\sigma'}}{\omega+\mu-\varepsilon_f-\Delta-U},
\end{eqnarray}
which gives the lower and upper Hubbard bands in the atomic limit. However, this approximation can not explain the Kondo physics because it can not produce the Kondo peak at the Fermi level. Consequently, decoupling schemes beyond the equivalent of the Hubbard-I approximation are needed in order to describe the metallic states and explain the Mott metal-insulator transition.

In actual materials calculations, the choice of the decoupling scheme is mainly decided by considering two factors: the accuracy and the cpu time consumption. The accuracy decides whether the exactly included interactions can be sufficient to describe a system. For example, for a complete insulator, the equivalent Hubbard-I approximation is already enough and will save a lot of cpu time over the use of the higher order decoupling schemes. But for metal and Kondo physics, the decoupling scheme beyond the Hubbard-I has to be employed. Moreover, by comparing different order of decoupling schemes, it is easy to examine the contribution of the involved interactions, which may provide understanding of the underlying many-body physics.

Now we return to our studied multi-orbital problem.
 For simplicity, here we first treat the on-site inter-orbital Coulomb interaction with a mean field approximation which is valid when the inter-orbital fluctuations are weak in the system, i.e., we use the corresponding decoupling
\begin{eqnarray}
&&\ll\hat{n}_{l\sigma}f_{m\sigma};f^{\dag}_{m\sigma}\gg \, \approx \bar{n}_{l\sigma}\ll~f_{m\sigma};f^{\dag}_{m\sigma}\gg,\nonumber\\
&&\ll\hat{n}_{l\sigma'}f_{m\sigma};f^{\dag}_{m\sigma}\gg \, \approx \bar{n}_{l\sigma'}\ll~f_{m\sigma};f^{\dag}_{m\sigma}\gg.
\end{eqnarray}
With this decoupling the on-site inter-orbital fluctuations are neglected, which may be a loss of some interesting information.
These on-site inter-orbital fluctuations will be taken into account in a forthcoming work. 

According to \eref{eq:6} and \eref{eq:7}, now the hybridization function $\Delta$ in single orbital case will change to the hybridization function $\Delta_{m\sigma}$ for the $m$-th orbital,
\begin{eqnarray}
\Delta_{m\sigma}&=&\Delta_{mm\sigma}+\sum_{l,l\neq m} \Delta_{lm\sigma},\nonumber\\
\Delta_{mm\sigma}&=&\sum_k\frac{V_{mk\sigma}{\cal V}^{\ast}_{mmk\sigma}}{\omega+\mu-\varepsilon_{km\sigma}}\nonumber\\
\Delta_{lm\sigma}&=&\sum_k\frac{V_{mk\sigma}{\cal V}^{\ast}_{lmk\sigma}}{\omega+\mu-\varepsilon_{km\sigma}}\frac{\ll
f_{l\sigma};f^{\dag}_{m\sigma}\gg}{\ll
f_{m\sigma};f^{\dag}_{m\sigma}\gg},
\label{eq:75}
\end{eqnarray}
where the $\Delta_{mm\sigma}$ ($\Delta_{lm\sigma}$) are the on-site indirect identical orbital (inter-orbital) hybridizations, which relate to the inter-site diagonal (off-diagonal) hopping terms. First we shall not discuss the effect of $\Delta_{mm\sigma}$ and $\Delta_{lm\sigma}$, and use only the symbol of total hybridization $\Delta_{m\sigma}$ in solving the equations of motion. We will then discuss the parts of $\Delta_{m\sigma}$ later in this section along with the DMFT self-consistency conditions.

Thus \eref{eq:7} will be replaced with,
\begin{eqnarray}
\sum_kV_{mk\sigma}\ll c_{mk\sigma};f^{\dag}_{m\sigma}\gg=\Delta_{m\sigma}\ll f_{m\sigma};f^{\dag}_{m\sigma}\gg.
\end{eqnarray}

Moreover, because we will mainly concentrate on the on-site Coulomb interactions and inter-site hopping effects in \eref{eq:1}, we first neglect the on-site direct inter-orbital hopping terms, which will be studied in a forthcoming work.

Then we can obtain
\begin{eqnarray}
\big(\omega+\mu-\varepsilon_{fm\sigma}-\sum_{l,l\neq m}(U_{lm\sigma\sigma}\bar{n}_{l\sigma}+U_{lm\sigma\sigma'}\bar{n}_{l\sigma'})-\Delta_{m\sigma}\big)\times~~~&&\nonumber\\
\ll f_{m\sigma}^{~} ;f^{\dag}_{m\sigma}\gg
=1+U^{mm}_{\it eff}\ll\hat{n}_{m\sigma'}f_{m\sigma}^{~};f^{\dag}_{m\sigma}\gg,~&&
\label{eq:8}
\end{eqnarray}
where
\begin{equation}
\bar{n}_{l\sigma} = <\hat{n}_{l\sigma}>=-\frac{1}{\pi}\int\ d\omega f(\omega-\mu)\imag{\ll
f_{l\sigma}^{~};f^{\dag}_{l\sigma}\gg}
\end{equation}
is occupation number in the $l$-th orbital and
$f(\omega)$ is the Fermi distribution function.

We can see that the existence of electrons in other orbitals has promoted the initial orbital level in single orbital case $\varepsilon_{fm\sigma}$ to an orbital level in multi-orbital case,
\begin{equation}
E_{fm\sigma}=\varepsilon_{fm\sigma}+\sum_{{l}, \,{l\neq m}}\big(U_{lm\sigma\sigma}\bar{n}_{l\sigma}+U_{lm\sigma\sigma'}\bar{n}_{l\sigma'}\big).
\label{eq:85}
\end{equation}
In real calculations, due to the partial filling and the fact that this inter-orbital Coulomb interaction should act on the ``charge center" of the $m$-th orbital, this will be modified as
\begin{eqnarray}
E_{fm\sigma} &= &\varepsilon_{fm\sigma} +(1-\bar{n}_{m\sigma'}) \times \nonumber \\
& & \sum_{{l},\,{l\neq m}}\big(U_{lm\sigma\sigma}\bar{n}_{l\sigma}\bar{n}_{m\sigma}+U_{lm\sigma\sigma'}\bar{n}_{l\sigma'}\bar{n}_{m\sigma}\big),\label{eq:86}
\end{eqnarray}
an initial value of which can be extracted directly from density functional theory calculations, in contrast to $\varepsilon_{fm\sigma}$, when we study a multi-orbital system. Afterwards, the orbital levels will be corrected iteratively by the occupations obtained in the DMFT calculation in each iteration. Moreover, we note that, for a system with fixed filling, the Fermi level should also shift accordingly.

Due to the existence of charges in other orbitals, the energy to add one electron into the $m$-th orbital has also changed from $U_{mm}$ to an effective value
\begin{eqnarray}
U^{mm}_{\it eff}=U_{mm}+\sum_{l,l\neq m}\big(U_{lm\sigma\sigma}\bar{n}_{l\sigma}+U_{lm\sigma\sigma'}\bar{n}_{l\sigma'}\big).
\end{eqnarray}

For higher order Green's functions, we obtain the equations of motion

\begin{eqnarray}
\fl (\omega+\mu-\varepsilon_{fm\sigma})G^{m}_{nf}
&=&\bar{n}_{m\sigma'}+U_{mm}\ll~\hat{n}_{m\sigma'}f_{m\sigma};f^{\dag}_{m\sigma}\gg\nonumber\\
&+&\sum_l\big(U_{lm\sigma\sigma}\ll~\hat{n}_{l\sigma}\hat{n}_{m\sigma'}f_{m\sigma};f^{\dag}_{m\sigma}\gg+U_{lm\sigma'\sigma}\ll
\hat{n}_{l\sigma'}\hat{n}_{m\sigma'}f_{m\sigma};f^{\dag}_{m\sigma}\gg\big)\nonumber\\
&+&\sum_k\big(-V^{\ast}_{mk\sigma'}\ll
c^{\dag}_{mk\sigma'}f_{m\sigma'}f_{m\sigma};f^{\dag}_{m\sigma}\gg+V_{mk\sigma}\ll
\hat{n}_{m\sigma'}c_{mk\sigma};f^{\dag}_{m\sigma}\gg\nonumber\\
& &~~~~~~+ V_{mk\sigma'}\ll
f^{\dag}_{m\sigma'}c_{mk\sigma'}f_{m\sigma};f^{\dag}_{m\sigma}\gg\big) ,
\label{eq:9}
\end{eqnarray}
\begin{eqnarray}
\fl (\omega+\mu-\varepsilon_{km\sigma})G^{m}_{nc}
&=&{\cal V}^{\ast}_{mmk\sigma}\ll~\hat{n}_{m\sigma'}f_{m\sigma};f^{\dag}_{m\sigma}\gg
+\sum_{k'}\big(V^{\ast}_{mk'\sigma'}\ll~f^{\dag}_{m\sigma'}c_{mk'\sigma'}c_{mk\sigma};f^{\dag}_{m\sigma}\gg \nonumber\\
&-&V_{mk'\sigma'}\ll~c^{\dag}_{mk'\sigma'}f_{m\sigma'}c_{mk\sigma};f^{\dag}_{m\sigma}\gg\big)
+\sum_{l,l\neq m} {\cal V}^{\ast}_{lmk\sigma}\ll~\hat{n}_{m\sigma'}f_{l\sigma};f^{\dag}_{m\sigma}\gg ,\nonumber\\
\label{eq:10}
\end{eqnarray}
\begin{eqnarray}
\fl (\omega+\mu+\varepsilon_{fm\sigma'}-\varepsilon_{km\sigma'}-\varepsilon_{fm\sigma})G^{m}_{fcf}
&=&\langle f^{\dag}_{m\sigma'}c_{mk\sigma'}\rangle +{\cal V}^{\ast}_{mmk\sigma'}\ll~\hat{n}_{m\sigma'}f_{m\sigma};f^{\dag}_{m\sigma}\gg\nonumber \\
&&+\sum_{k'}\big(V^{\ast}_{mk'\sigma'}\ll~f^{\dag}_{m\sigma'}c_{mk\sigma'}c_{mk'\sigma};f^{\dag}_{m\sigma}\gg\nonumber\\
&&-V_{mk'\sigma'}\ll~c^{\dag}_{mk'\sigma'}c_{mk\sigma'}f_{m\sigma};f^{\dag}_{m\sigma}\gg\big)\nonumber\\
&&+\sum_{l,l\neq m} {\cal V}^{\ast}_{lmk\sigma'}\ll~f^{\dag}_{m\sigma'}f_{l\sigma'}f_{m\sigma};f^{\dag}_{m\sigma}\gg ,
\label{eq:11}
\end{eqnarray}
\begin{eqnarray}
\fl (\omega+\mu+\varepsilon_{km\sigma'}-\varepsilon_{fm\sigma'}-\varepsilon_{fm\sigma})G^{m}_{cff}
&=&\langle c^{\dag}_{mk\sigma'}f_{m\sigma'}\rangle +U_{mm}\ll~c^{\dag}_{mk\sigma'}f_{m\sigma'}f_{m\sigma};f^{\dag}_{m\sigma}\gg \nonumber \\
&&+ 2\sum_l\big(U_{lm\sigma\sigma}\ll~\hat{n}_{l\sigma}c^{\dag}_{mk\sigma'}f_{m\sigma'}f_{m\sigma};f^{\dag}_{m\sigma}\gg \nonumber \\
&&~~~~~~~~~~+U_{lm\sigma'\sigma}\ll
\hat{n}_{l\sigma'}c^{\dag}_{mk\sigma'}f_{m\sigma'}f_{m\sigma};f^{\dag}_{m\sigma}\gg \big)\nonumber\\
&&-{\cal V}_{mmk\sigma'}\ll~\hat{n}_{m\sigma'}f_{m\sigma};f^{\dag}_{m\sigma}\gg\nonumber\\
&&+\sum_{k'}\big(V^{\ast}_{mk'\sigma'}\ll~c^{\dag}_{mk\sigma'}c_{mk'\sigma'}f_{m\sigma};f^{\dag}_{m\sigma}\gg\nonumber\\
&&+V^{\ast}_{mk'\sigma}\ll~c^{\dag}_{mk\sigma'}f_{m\sigma'}c_{mk'\sigma};f^{\dag}_{m\sigma}\gg\big)\nonumber\\
&&-\sum_{l,l\neq m} {\cal V}_{lmk\sigma'}\ll~f^{\dag}_{l\sigma'}f_{m\sigma'}f_{m\sigma};f^{\dag}_{m\sigma}\gg,
\label{eq:12}
\end{eqnarray}
where we have used the following abbreviations on the left hand side (LHS) of the equations:
\begin{eqnarray} 
G^{m}_{nf}&=&\ll~\hat{n}_{m\sigma'}f_{m\sigma};f^{\dag}_{m\sigma}\gg,\nonumber\\
G^{m}_{nc}&=&\ll~\hat{n}_{m\sigma'}c_{mk\sigma};f^{\dag}_{m\sigma}\gg,\nonumber\\
G^{m}_{fcf}&=&\ll~f^{\dag}_{m\sigma'}c_{mk\sigma'}f_{m\sigma};f^{\dag}_{m\sigma}\gg,\nonumber\\
G^{m}_{cff}&=&\ll~c^{\dag}_{mk\sigma'}f_{m\sigma'}f_{m\sigma};f^{\dag}_{m\sigma}\gg.\nonumber
\end{eqnarray}

For the three-particle Green's functions, we introduce according to the mean field approximation for inter-band Coulomb interactions,
\begin{eqnarray}
\ll~\hat{n}_{l\sigma}\hat{n}_{m\sigma'}f_{m\sigma}^{~};f^{\dag}_{m\sigma}\gg \,\approx\bar{n}_{l\sigma}\ll~\hat{n}_{m\sigma'}f_{m\sigma}^{~};f^{\dag}_{m\sigma}\gg,~\\
\ll~\hat{n}_{l\sigma'}\hat{n}_{m\sigma'}f_{m\sigma}^{~};f^{\dag}_{m\sigma}\gg \, \approx\bar{n}_{l\sigma'}\ll~\hat{n}_{m\sigma'}f_{m\sigma}^{~};f^{\dag}_{m\sigma}\gg,~
\end{eqnarray}
the effect of this is to increase the impurity position in the same way as in \eref{eq:85} when it moves to the LHS in Eqs.\ \eref{eq:9}-\eref{eq:12} as pre-existed charge in other orbitals. For the same reason
 the Coulomb interaction strength $U_{mm}$ will change to $U^{mm}_{\it eff}$. Moreover, if there is no external field existing, the energy levels for different spins in the same orbital should have $E_{fm\sigma'}=E_{fm\sigma}$.

For the Green's functions involving an odd number of operators of the $m$-th orbital, we follow our previous assumption in \eref{eq:75} and treat them as zero.
For the Green's functions involving only operators of the $m$-th orbital, we use the decoupling applied in \cite{FZJ09}.

Solving now the closed set of equations, we obtain the single particle Green's functions

\begin{eqnarray}
\fl \ll f_{m\sigma}^{~};f^{\dag}_{m\sigma}\gg=\frac{1+\frac{U^{mm}_{\it eff}}{\omega+\mu-E_{fm\sigma}-U^{mm}_{\it eff}-\Delta_{m\sigma}-\Delta_{m\sigma1}-\tilde{\Delta}_{m\sigma}} \bigg(\bar{n}_{m\sigma'}+I_{1}\bigg)}
{\omega+\mu-E_{fm\sigma}-\Delta_{m\sigma}-\frac{U^{mm}_{\it eff}}{\omega+\mu-E_{fm\sigma}-U^{mm}_{\it eff}-\Delta_{m\sigma}-\Delta_{m\sigma1}-\tilde{\Delta}_{m\sigma}} \big(\Delta_{m\sigma}\cdot I_1+I_2\big)},\label{eq:singleparticleGF}
\end{eqnarray}
where
\begin{eqnarray}
\Delta_{m\sigma1}&=&\sum_k\frac{{\cal V}^{\ast}_{mmk\sigma'}V_{mk\sigma'}^{~}}{\omega+\mu+E_{fm\sigma'}-E_{fm\sigma}-\varepsilon_{km\sigma'}}\nonumber\\
&&+\sum_{kl,l\neq m}\frac{{\cal V}^{\ast}_{lmk\sigma'}V_{mk\sigma'}^{~}}{\omega+\mu+E_{fm\sigma'}-E_{fl\sigma}-\varepsilon_{km\sigma'}}\frac{\ll
f_{l\sigma'};f^{\dag}_{m\sigma'}\gg}{\ll
f_{m\sigma'};f^{\dag}_{m\sigma'}\gg},
\end{eqnarray}
\begin{eqnarray}
\tilde{\Delta}_{m\sigma}&=&\sum_k\frac{V^{\ast}_{mk\sigma'}{\cal V}_{mmk\sigma'}^{~}}{\omega+\mu+\varepsilon_{km\sigma'}-E_{fm\sigma'}-E_{fm\sigma}-U^{mmb}_{\it eff}}\nonumber\\
&&+\sum_{kl,l\neq m}\frac{V^{\ast}_{mk\sigma'}{\cal V}_{lmk\sigma'}^{~}}{\omega+\mu+\varepsilon_{km\sigma'}-E_{fm\sigma'}-E_{fl\sigma}-U^{mmb}_{\it eff}}
\frac{\ll
f_{l\sigma'};f^{\dag}_{m\sigma'}\gg}{\ll
f_{m\sigma'};f^{\dag}_{m\sigma'}\gg},
\end{eqnarray}
\begin{eqnarray}
\fl I_1&=&\sum_{k}\big(\frac{V^{\ast}_{mk\sigma'}\langle f^{\dag}_{m\sigma'}c_{mk\sigma'}^{~}\rangle}{\omega+\mu+E_{fm\sigma'}-E_{fm\sigma}-\varepsilon_{km\sigma}}-\frac{V_{mk\sigma'}\langle c^{\dag}_{mk\sigma'}f_{m\sigma'}^{~}\rangle}{\omega+\mu+\varepsilon_{km\sigma}-E_{fm\sigma'}-E_{fm\sigma}-U^{mmb}_{\it eff}}\big),\label{eq:90000}
\end{eqnarray}
\begin{eqnarray}
\fl I_2&=&-\sum_{kk'}\big(\frac{V^{\ast}_{mk\sigma'}V_{mk\sigma'}^{~}\langle c^{\dag}_{mk'\sigma'}c_{mk\sigma'}^{~}\rangle}{\omega+\mu+E_{fm\sigma'}-E_{fm\sigma}-\varepsilon_{km\sigma}}
+\frac{V^{\ast}_{mk\sigma'}V_{mk\sigma'}^{~}\langle c^{\dag}_{mk\sigma'}c_{mk'\sigma'}^{~}\rangle}{\omega+\mu+\varepsilon_{km\sigma}-E_{fm\sigma'}-E_{fm\sigma}-U^{mmb}_{\it eff}}\big),\label{eq:90001}
\end{eqnarray}
\begin{eqnarray}
U^{mmb}_{\it eff}=U_{mm}+2\sum_{l,l\neq m}\big(U_{lm\sigma\sigma}\bar{n}_{l\sigma}+U_{lm\sigma\sigma'}\bar{n}_{l\sigma'}\big).\nonumber
\end{eqnarray}
and furthermore we will use the Hermitian relations
\begin{eqnarray}
\langle f^{\dag}_{m\sigma'}c_{mk\sigma'}^{~}\rangle=\langle c^{\dag}_{mk\sigma'}f_{m\sigma'}^{~}\rangle,\qquad \langle c^{\dag}_{mk'\sigma'}c_{mk\sigma'}^{~}\rangle=\langle c^{\dag}_{mk\sigma'}c_{mk'\sigma'}^{~}\rangle.\nonumber
\end{eqnarray}

In \eref{eq:singleparticleGF}, if $I_1, I_2$ are taken as zero, it turns to a multi-orbital Hubbard-I method with the mean field approximation to the inter-orbital Coulomb interactions. , which will be reliable at an atomic limit, i.e., for completely insulating states.

Let us discuss now on the hybridizations.
If we neglect the hybridization between different orbitals (i.e., neglect the inter-site inter-orbital hopping in \eref{eq:1}, and then the model Hamiltonian of \eref{eq:1} will return to the Hamiltonian in \cite{OSMT1}), the above formula can be simplified by setting ${\cal V}^{\ast}_{lmk\sigma}=0$. In this sense, the approximation is similar to the single orbital case with arbitrary spin-orbital degeneracy $N$, where the degenerated electrons only hybridize with electrons having identical spin and orbital indices in the bath. The main difference between the $N$-fold single orbital case and the above multi-orbital impurity solver is that the latter one is generalized to $N/2$-orbital system with different orbital levels, band widths and varied inter-orbital and intra-orbital Coulomb interaction strengths for different orbitals. For this case, the DMFT self-consistency condition
on the Bethe lattice should be
\begin{eqnarray}
\Delta_{m\sigma}=t^2_m \ll f_{m\sigma}^{~ };f^{\dag}_{m\sigma}\gg ,
\end{eqnarray}
just like the DMFT self-consistency condition for a single orbital Hubbard model with SU($N$) symmetry.

Now let us study the equations in a further step as we mentioned above.
If we take into account the indirect hybridizations between different orbitals, we will obtain two new equations of motion as follows,
\begin{eqnarray}
\fl (\omega+\mu-\varepsilon_{fl\sigma})\ll
f_{l\sigma}^{~};f^{\dag}_{m\sigma}\gg
&=&\langle [f_{l\sigma}^{~},f^{\dag}_{m\sigma}]_+\rangle+U_{ll}\ll\hat{n}_{l\sigma'}f_{l\sigma}^{~};f^{\dag}_{m\sigma}\gg\nonumber\\
&+&\sum_{l',l'\neq l}\big(U_{ll'\sigma\sigma}\ll\hat{n}_{l'\sigma}f_{l\sigma}^{~};f^{\dag}_{m\sigma}\gg\nonumber\\
&&\qquad+U_{ll'\sigma\sigma'}\ll\hat{n}_{l'\sigma'}f_{l\sigma};f^{\dag}_{m\sigma}\gg\big)\nonumber\\
&+&\sum_{k}V_{lk\sigma}\ll c_{lk\sigma}^{~};f^{\dag}_{m\sigma}\gg
,\label{eq:71}\\
\fl (\omega+\mu-\varepsilon_{kl\sigma})\ll
c_{lk\sigma}^{~};f^{\dag}_{m\sigma}\gg&=&
{\cal V}^{\ast}_{lmk\sigma}\ll
f_{m\sigma}^{~};f^{\dag}_{m\sigma}\gg+{\cal V}^{\ast}_{llk\sigma}\ll f_{l\sigma}^{~};f^{\dag}_{m\sigma}\gg_{l\neq m}\nonumber\\
&+&\sum_{l'}{\cal V}^{\ast}_{ll'k\sigma}\ll f_{l'\sigma}^{~};f^{\dag}_{m\sigma}\gg_{\stackrel{l'\neq l}{l'\neq m}}
.\label{eq:72}
\end{eqnarray}
Within the mean field approximation, we can then obtain after charge correction (see \eref{eq:86})
\begin{eqnarray}
&&\fl \ll f_{l\sigma}^{~};f^{\dag}_{m\sigma}\gg\nonumber\\
&&\fl ~~~=\frac{\langle [f_{l\sigma}^{~},f^{\dag}_{m\sigma}]_+\rangle+\stackrel{\sum}{_{l',l'\neq l}}\big(\sum_{k}\frac{{\cal V}^{\ast}_{ll'k\sigma}V_{lk\sigma}^{~}}{\omega+\mu-\varepsilon_{kl\sigma}}
\big)\ll f_{l'\sigma}^{~};f^{\dag}_{m\sigma}\gg+\frac{U^{ll}_{\it eff}}{\omega+\mu-E_{fl\sigma}-U^{ll}_{\it eff}-\Delta_{l\sigma}-\Delta_{l\sigma1}-\tilde{\Delta}_{l\sigma}}\bar{n}_{l\sigma'}}
{\omega+\mu-E_{fl\sigma}-\Delta_{l\sigma}-\frac{U^{ll}_{\it eff}}{\omega+\mu-E_{fl\sigma}-U^{ll}_{\it eff}-\Delta_{l\sigma}-\Delta_{l\sigma1}-\tilde{\Delta}_{l\sigma}}\Delta_{l\sigma}\cdot \sum_k\frac{V^{\ast}_{lk'\sigma'}V_{lk\sigma'}^{~}\langle c^{\dag}_{lk'\sigma'}c_{lk\sigma'}^{~}\rangle}{\omega+\mu-\varepsilon_{kl\sigma}}},\label{eq:10000}
\end{eqnarray}

where we have taken the two correlation functions as
\begin{eqnarray}
\langle f^{\dag}_{l\sigma'}c_{lk\sigma'}^{~}\rangle
=0 ,\qquad
\end{eqnarray}
\begin{eqnarray}
\langle c^{\dag}_{lk'\sigma'}c_{lk\sigma'}^{~}\rangle
=-\frac{1}{\pi}\int
d\omega'f(\omega'-\mu)\imag\frac{\delta_{kk'}}{\omega'+\mu-\varepsilon_{k'l\sigma'}}.
\end{eqnarray}

\begin{figure}[tb!]
\begin{center}
 \includegraphics[angle=0,width=6cm]{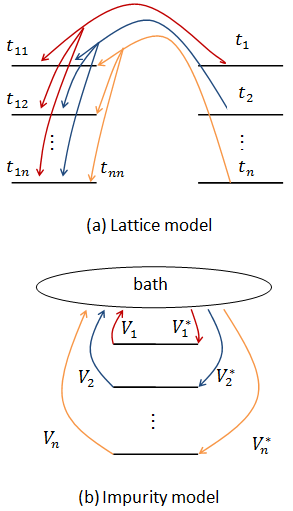}%
\end{center}
\caption{(Color online) Illustration of inter-site multi-orbital hopping in a Hubbard type lattice model (a) and hybridizations in a mapped impurity model (b). In the Hubbard type lattice model, we only show the hoppings in one direction. If neighbouring sites are identical, the hopping in opposite direction will be the same.
 \label{figure1}}
\end{figure}

It is a general way to exactly solve the hybridization functions $\Delta_{m\sigma}$ by including \eref{eq:10000} which will make the equations fully closed for the impurity solver. However the set of equations \eref{eq:10000} only relate to the hybridization functions. If one has  another method to obtain the hybridization functions, the set of equations from \eref{eq:singleparticleGF} to \eref{eq:90001} will be closed where $\Delta_{m\sigma1}$ and $\tilde{\Delta}_{m\sigma}$ can be obtained from $\Delta_{m\sigma}$ with interpolation, and the correlation functions in \eref{eq:90000} and \eref{eq:90001} are calculated with the obtained hybridization functions and single particle Green's functions through the spectral theorem. Fortunately, we are now studying a many body system using DMFT. Therefore, the hybridization functions $\Delta_{m\sigma}$ do not need to be exactly solved in the impurity solver because in the hybridization functions in DMFT will be eventually determined by the DMFT self-consistency conditions in each DMFT iteration. The exact solution of the hybridization functions will however be needed when only the single impurity Anderson model is studied. For DMFT, even if we obtained the exact hybridization functions of the impurity, these are only used in the first iteration (guess of the initial GFs) and then replaced by new hybridization functions determined by DMFT self-consistency conditions from the second DMFT iteration, while using an exact hybridization function or an approximate hybridization function in \eref{eq:singleparticleGF} in the first DMFT iteration makes little difference for the final result. Moreover, for orthogonal orbitals, the off-diagonal single particle GFs need not be calculated. Principally in DMFT for a general lattice, the hybridization function should be determined self-consistently together with the Green's function and self-energy by summing over $k$ throughout the whole Brillouin zone. However, for the Bethe lattice, it will be much easier because it is simply determined by the inter-site hopping. Hence, we use the Bethe lattice here as an example. Assuming the probability of electrons outgoing from one orbital to a neighbour site is identical to the probability of electrons coming from the neighbour site into the orbital with identical orbital index. This is reasonable for an equilibrium system. Then, for the Bethe lattice, the DMFT self-consistency condition approximately is
\begin{eqnarray}
\fl \Delta_{m\sigma}
&=&\Delta_{mm\sigma}+\sum_{l,l\neq m}\Delta_{lm\sigma}
=t_m^2\frac{t^2_m}{t^2_{tot}}\ll f_{m\sigma}^{~};f^{\dag}_{m\sigma}\gg
+\sum_{l,l\neq m} t^2_m \frac{t^2_l}{t^2_{tot}}\ll f_{l\sigma}^{~};f^{\dag}_{l\sigma}\gg,
\end{eqnarray}
where $t_m$ is the total probability for the $i$-site electron with spin $\sigma$ in the $m$-th orbital hopping to the $j$-site, $t_{tot}=\sum_{m}t_m$ is the total probability for all $i$-site electrons with spin $\sigma$ in all localized orbitals hopping to $j$-site, and $t_{ml}=\frac{t_m t_l}{t_{tot}}$ is the probability for the $i$-site electron with spin $\sigma$ in the $m$-th orbital hopping to the $l$-th orbital of the $j$-site, as illustrated in Figure\ \ref{figure1},
and so as in the opposite direction. In this case the numbers of outgoing and incoming hopping electrons are identical in each orbital so that the system is in an equilibrium state.

Moreover, when one orbital is not occupied, e.g., $\bar{n}_{l\sigma}=\bar{n}_{l\sigma'}=0$, the equations for the $N$-orbital system automatically turn to those for $(N-1)$ orbitals.

For another lattice, these set of hybridization functions $\Delta_{m\sigma}$ will be calculated by an integral over $k$ space for conduction electrons, then be put into the equations to calculate the single particle Green's functions in each orbital.

At the end of this section, we would like to give a discussion of the spin-flip term $f^{\dag}_{im\sigma'}f^{\dag}_{il\sigma}f_{il\sigma'}f_{im\sigma}$ and the pair-hopping term $f^{\dag}_{im\sigma'}f^{\dag}_{im\sigma}f_{il\sigma'}f_{il\sigma}$ which are studied in some recent literatures (see, e.g.,~\cite{CFCT09,Yin11}), but are not studied in this paper. One background for this is that contemporary LDA+DMFT calculations are mostly implemented to achieve the correct charge in the  strongly localized orbitals and then calculate the correct potential for a subsequent LDA calculation so that the spin-flip exchange term is less important in this context, which can be recognized from the following. All the Green's functions are associated with some kind of physical interactions, e.g.,
the Green's function $\ll \hat{n}_{fl\sigma}f_{m\sigma};f^{\dag}_{m\sigma}\gg$ labels an inter-orbital fluctuation which corresponds to the inter-orbital Coulomb interaction between electrons with identical spin.
In turn, the spin flip term is associated with a Green's function $\ll f^{\dag}_{l\sigma'}f_{m\sigma'}f_{l\sigma};f^{\dag}_{m\sigma}\gg$.
To show the influences of the Coulomb interactions and the spin-flip term, we make a comparison: for the Green's function $\ll \hat{n}_{fl\sigma}f_{m\sigma};f^{\dag}_{m\sigma}\gg$, one has
\begin{eqnarray}
\langle [\hat{n}_{fl\sigma}f_{m\sigma},f^{\dag}_{m\sigma}]_+\rangle=\bar{n}_{fl\sigma},
\end{eqnarray}
while for the spin-flip associated Green's function $\ll f^{\dag}_{l\sigma'}f_{m\sigma'}f_{l\sigma};f^{\dag}_{m\sigma}\gg$,
\begin{eqnarray}
\langle [f^{\dag}_{l\sigma'}f_{m\sigma'}f_{l\sigma},f^{\dag}_{m\sigma}]_+\rangle=0.
\end{eqnarray}
The above averages $\langle ... \rangle$ are corresponding to their connection to {\it charge} which will determine the effective Coulomb interaction strengths, the relative position of the two Hubbard bands, and the total densities of states in each of the Hubbard bands. One can note that the spin-flip term does not directly connect to charge for a system with the above mentioned anti-commutation relations for the operators, whereas the Coulomb interactions are associated with charge. It is estimated that the spin-flip associated Green's function will mainly contribute to the shape of the Hubbard bands which will show more micro structure on top of the Hubbard bands that are determined by the charge related Green's functions. Therefore, the Coulomb interactions are more important than the spin-flip term so that in this paper we first concentrate on the Coulomb interactions, while we have used the mean field approximation to treat the inter-orbital Coulomb interactions and have neglected the on-site inter-orbital fluctuations.  The spin-flip term will be studied in a higher-order decoupling scheme where the inter-orbital fluctuations associated with the inter-orbital Coulomb interactions will be fully taken into account. As for the pair-hopping term, this requires that the electron's starting orbital will be fully occupied and the destination orbital is empty. This physical image reflects a rare case and corresponds to the excited states. With the same reason as for the spin-flip term, we have neglected to study the pair-hopping term in this order of the decoupling scheme.

\section{Results and discussion\label{sect3}}
\begin{figure}[tb!]
\begin{center}
\includegraphics[angle=-90,width=7.5cm]{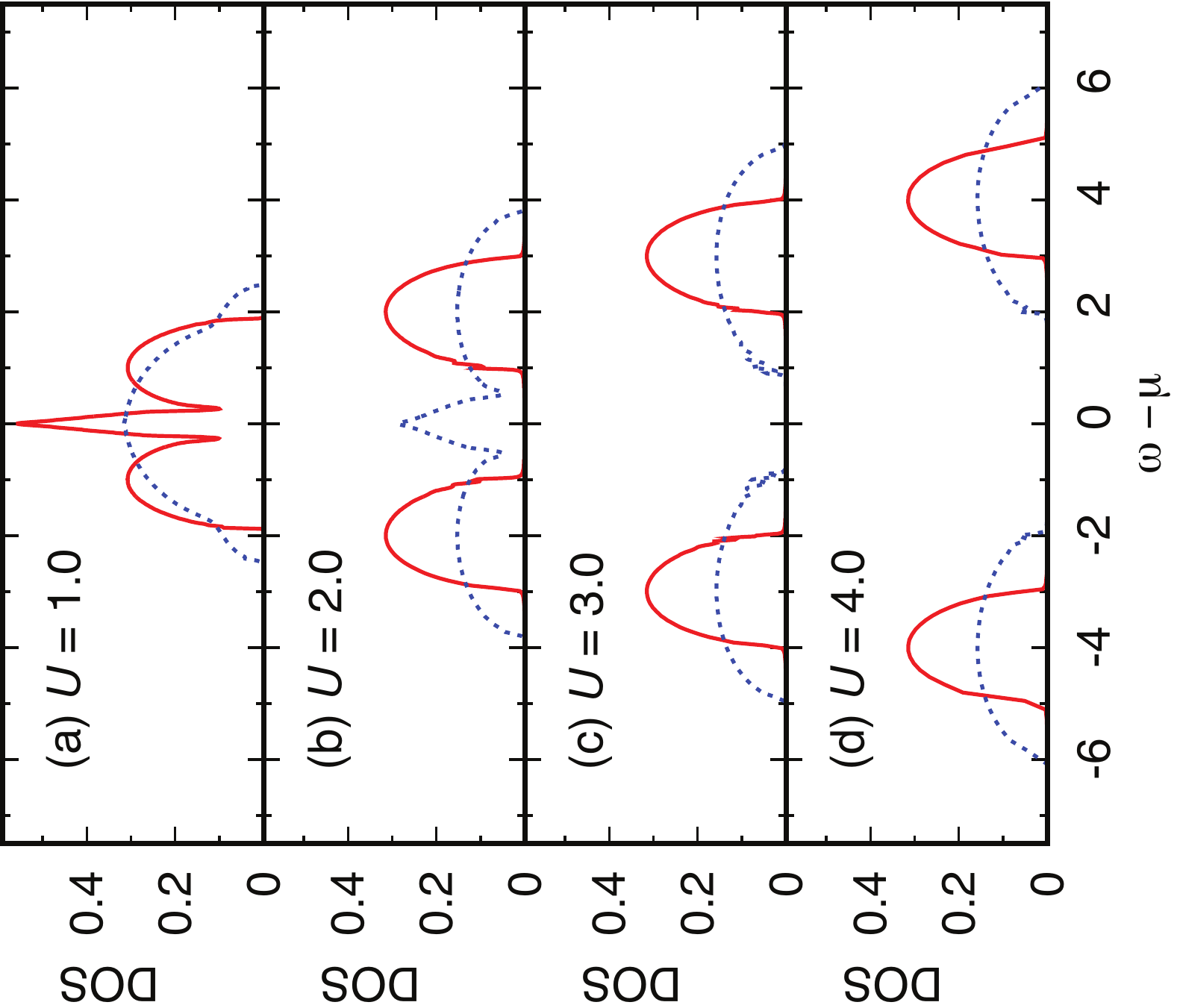}
\end{center}
\caption{(Color online) Quasiparticle densities of states for the halffilled two-orbital system on the Bethe lattice in case of neglecting inter-orbital hybridizations. The DOS are computed with half bandwidths $D_2=2D_1=2$, Hund's rule coupling constant $J=0$ and at temperature $T=0.01$. The red solid line gives the DOS of the narrow orbital and the blue dashed line gives that of the wide orbital.
\label{figure2}}
\end{figure}

Using the above derived MO-EOM impurity solver, we have investigated a two-orbital system with the two-orbital Hubbard model.
Here we first present calculated results for the case where the inter-orbital hybridization is neglected. In this case, our model Hamiltonian in \eref
{eq:1} will return to the frequently studied Hamiltonian as shown in \cite{OSMT1}. When the inter-orbital and intra-orbital Coulomb interaction strengths
are identical, i.e., $U_{11}=U_{22}=U_{12\sigma\sigma'}=U_{12\sigma\sigma}$ and the Hund's rule coupling constant $J=0$, we obtained the densities of
states (DOS) for the halffilled system, as shown in Figure~\ref{figure2}, where $\mu$ is the chemical potential and we have furthermore assumed
that the two occupied orbital levels have the relation $E_1=E_2$. We can see that the two orbitals have different broadening for the Hubbard bands. With increasing $U$, the system changes from metallic states to insulating states. However, the metal-insulator transitions (MIT) occur at different $U$, not shown in the plot. The critical value of $U$ for the two orbitals are correspondingly $U=U_{c1}\approx 1.5$ and $U=U_{c2}\approx 2.5$ (in units of half the narrow bandwidth).

\begin{figure}[tb!]
\begin{center}
\includegraphics[angle=-90,width=7.5cm]{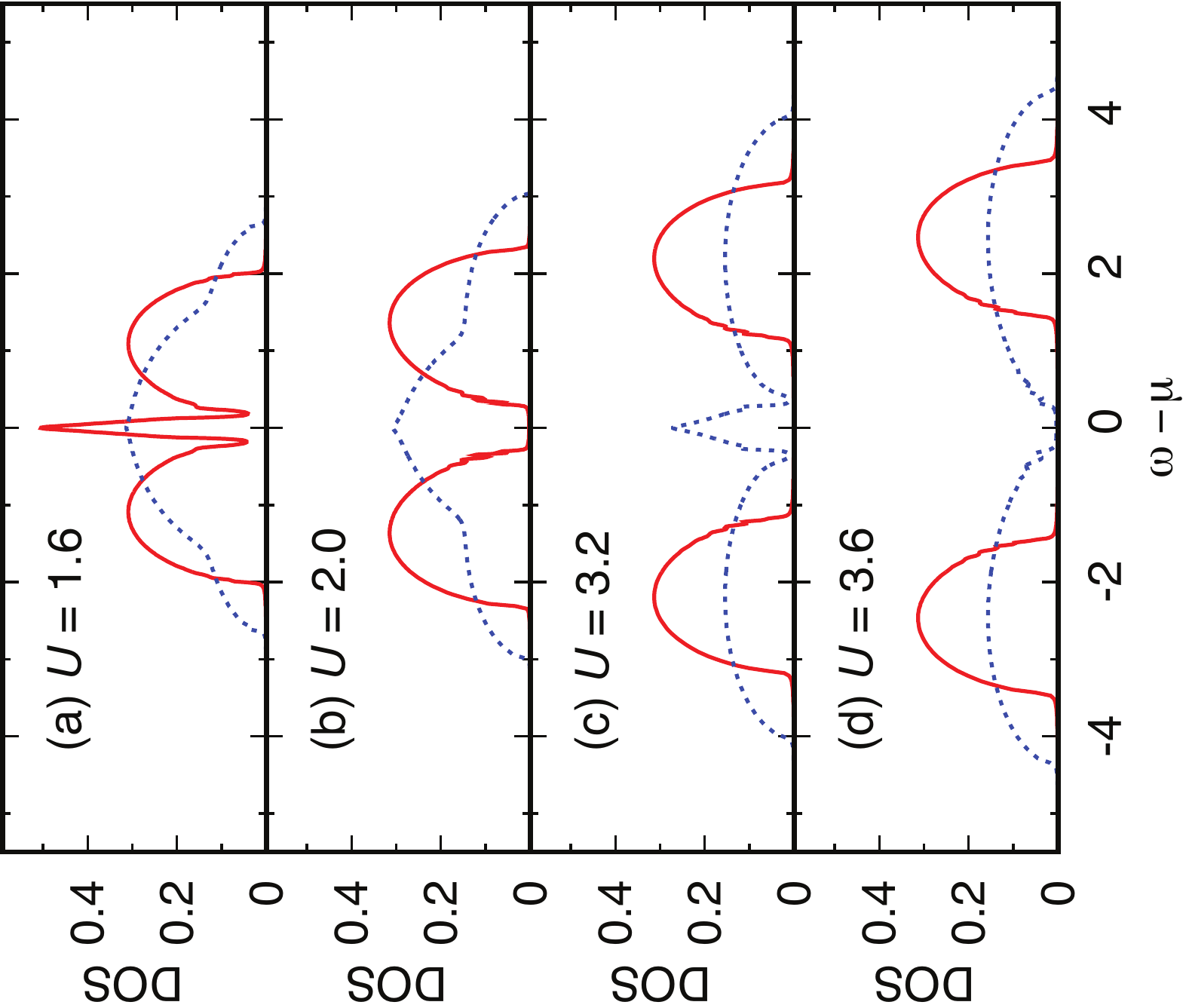}%
\end{center}
\caption{(Color online) Quasiparticle densities of states for the halffilled two-orbital system on the Bethe lattice in case of neglecting inter-orbital hybridizations, computed with half bandwidths $D_2=2D_1=2$, Hund's rule coupling constant $J=U/4$, and at temperature $T=0.01$.
The two orbitals are denoted by the dashed and full curves, as in Figure\ \ref{figure2}.
\label{figure3}}
\end{figure}
In Figure~\ref{figure3} we show the computed results for a non-zero $J$,  $J=U/4$. We still assume $U_{11}=U_{22}$, while the inter-orbital Coulomb interaction strength now changes to be $U_{12\sigma\sigma'}=U_{11}-2J$ and $U_{12\sigma\sigma}=U_{11}-3J$. Comparing to the identical $U$ case shown in Figure~\ref{figure2}, one recognizes that with the decrease of the inter-orbital Coulomb interaction strength, the effective $U^{mm}_{\it eff}$ decreases accordingly, which has influenced the MIT in each orbital. In this case, the critical value of $U$ for the two orbitals has correspondingly increased to $U=U_{c1}\approx 1.9$ and $U=U_{c2}\approx 3.5$, for the narrow and wide orbitals, respectively. In Figure~\ref{figure2} and Figure~\ref{figure3}, we have seen an orbital selective Mott transition (OSMT). But due to the neglect of inter-site inter-orbital hopping, here the two orbitals are only connected with the inter-orbital Coulomb interaction strengths and occupations. Comparing Figure~\ref{figure2} and Figure~\ref{figure3}, when the inter-orbital Coulomb interaction strengths decrease, the two Hubbard bands will come closer together so that the critical value of $U$ for the two orbitals will increase. Moreover, once the two orbitals have identical intra-orbital Coulomb interaction strengths and different bandwidth, the OSMT will always occur at the halffilling.

\begin{figure}[tb!]
\begin{center}
\includegraphics[angle=-90,width=8.5cm]{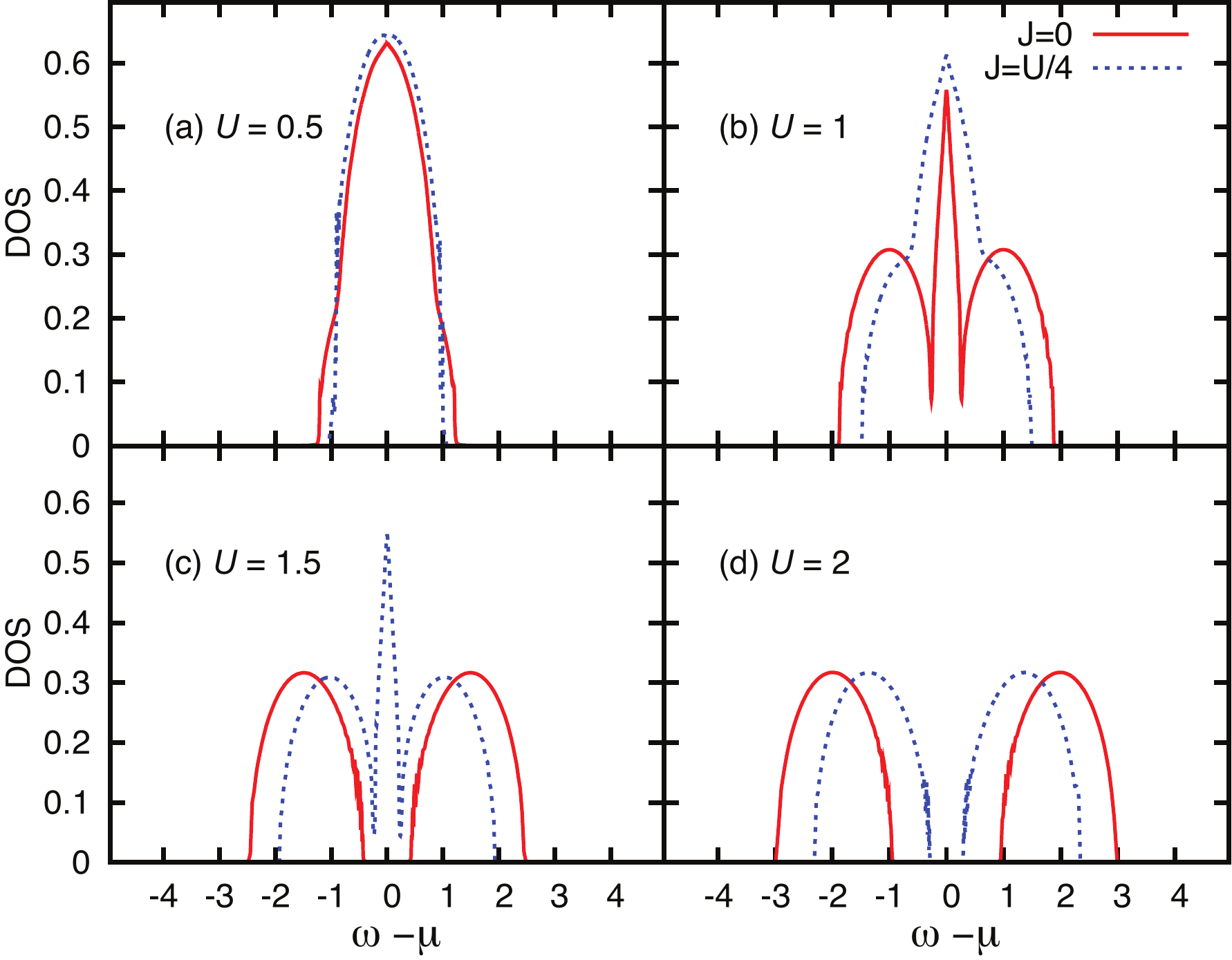}%
\end{center}
\caption{(Color online) Quasiparticle densities of states for the two-orbital system at the halffilling in case of including the effect of hopping between the orbitals. The DOS are computed with half bandwidths $D_1=D_2=1$ and at $T=0.01$.
 \label{figure4}}
\end{figure}

Next, we consider the two-orbital system taking now non-zero inter-orbital hybridizations into account.
Doing this, we have calculated the DOS for the corresponding two-orbital system at halffilling. First we show the obtained results in Figure~\ref{figure4} for the two-orbital system having identical bandwidths for the two orbitals. We can observe that the inclusion of the inter-orbital hybridizations has greatly changed the DOS. Besides the change of DOS with increasing $U$, we also give the comparison of the DOS at $J=0$ and $J=U/4$. When $J$ is not zero, the effective Coulomb interaction strength becomes smaller, which has increased the critical value of $U$ for the Mott transitions.

\begin{figure}[tb!]
\begin{center}
\includegraphics[angle=-90,width=8.5cm]{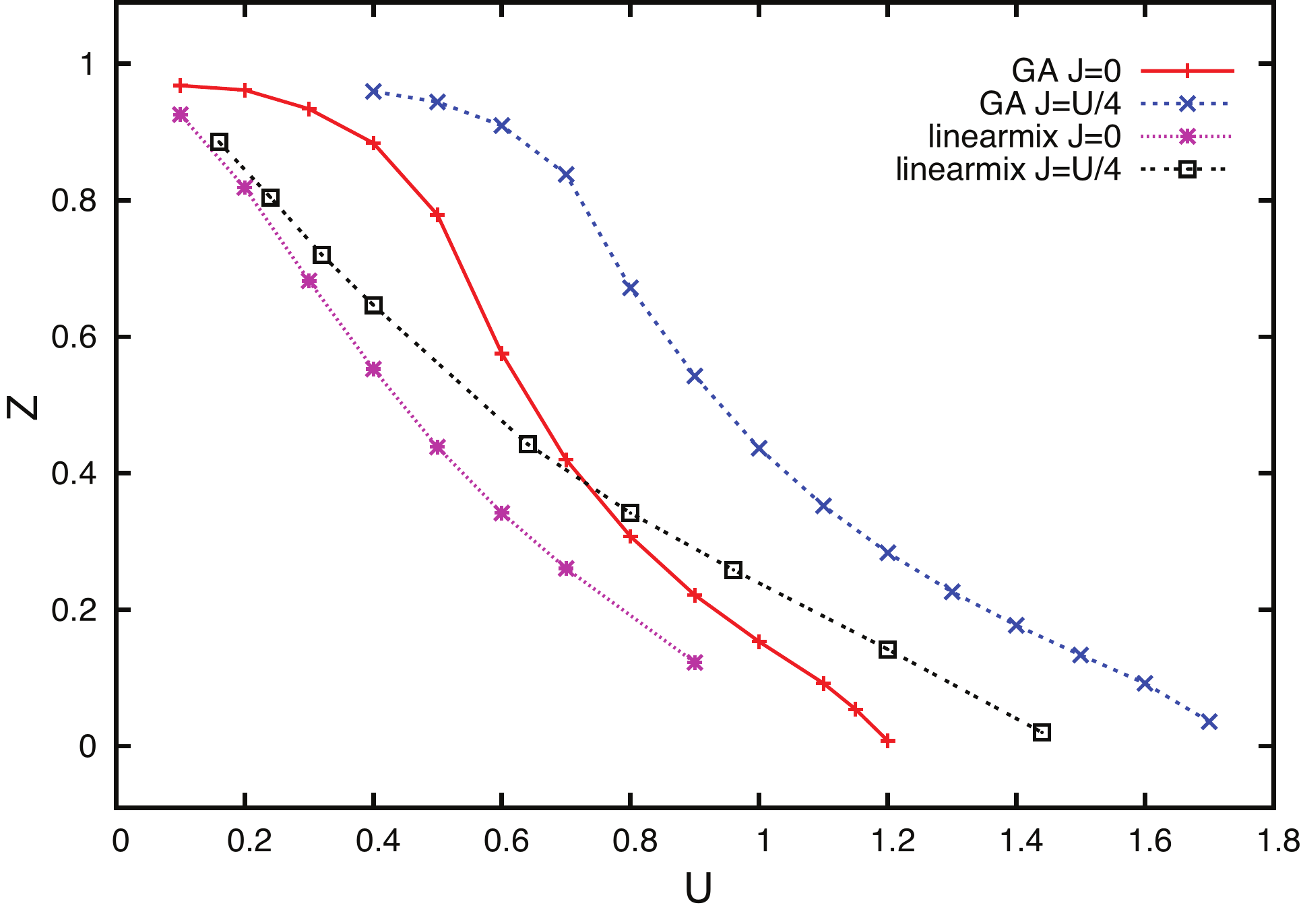}%
\end{center}
\caption{(Color online) Quasiparticle weight versus $U$ for the two-orbital system on the Bethe lattice at halffilling in the case of including hopping effect between the orbitals, computed with half bandwidths $D_1=D_2=1$ and at $T=0.01$.
 \label{figure5}}
\end{figure}
The quasiparticle weight, which is a characteristic quantity describing the strength of the correlation effect, is defined by
\begin{eqnarray} Z=\frac{m}{m^{\ast}}=\Big[1-\frac{\partial{\rm
Re}\Sigma(\omega)}{\partial\omega}\Big]^{-1}_{\omega\rightarrow 0},
\end{eqnarray}
where $\Sigma(\omega)$ is the self-energy, and $m$ and $m^{\star}$ the bare and enhanced mass, respectively. The quasiparticle weight has been computed as the function of $U$ at halffiling for the two-orbital system with identical bandwidths.
It is shown in Figure~\ref{figure5}. The abbreviation GA in the legend means that it is calculated with genetic algorithm (see \cite{FZJ09}), and ``linearmix" means that these data are calculated with the iterative method with linear mixing. The value of $Z$ decreases along with increasing the value  of $U$ until the critical $U_c$ for the Mott transition is reached.
However, due to the Lorentzian broadening applied in the iterative method with linear mixing, it has given minor different results and smaller critical values of $U$ for both $J=0$ and $J=U/4$ cases.

We mention here that all the model calculation results presented in this section, if not specified otherwise, are calculated with the genetic algorithm in order to efficiently solve the integral equations. Using instead the iterative method with linear mixing should have a minor difference on the computed DOS and critical value of $U$, but would nonetheless be less accurate and it will also cost more cpu time. This is so because the iterative method with linear mixing has to use a finite Lorentzian broadening in calculation of Green's functions on real frequencies. In calculations with the GA scheme, this Lorentzian broadening can be set to zero so that the GA scheme will give a numerically exact solution of EOMs. For a more detailed explanation, see \cite{phdthesis-feng}.

\begin{figure}[b!]
\begin{center}
\includegraphics[angle=-90,width=7.5cm]{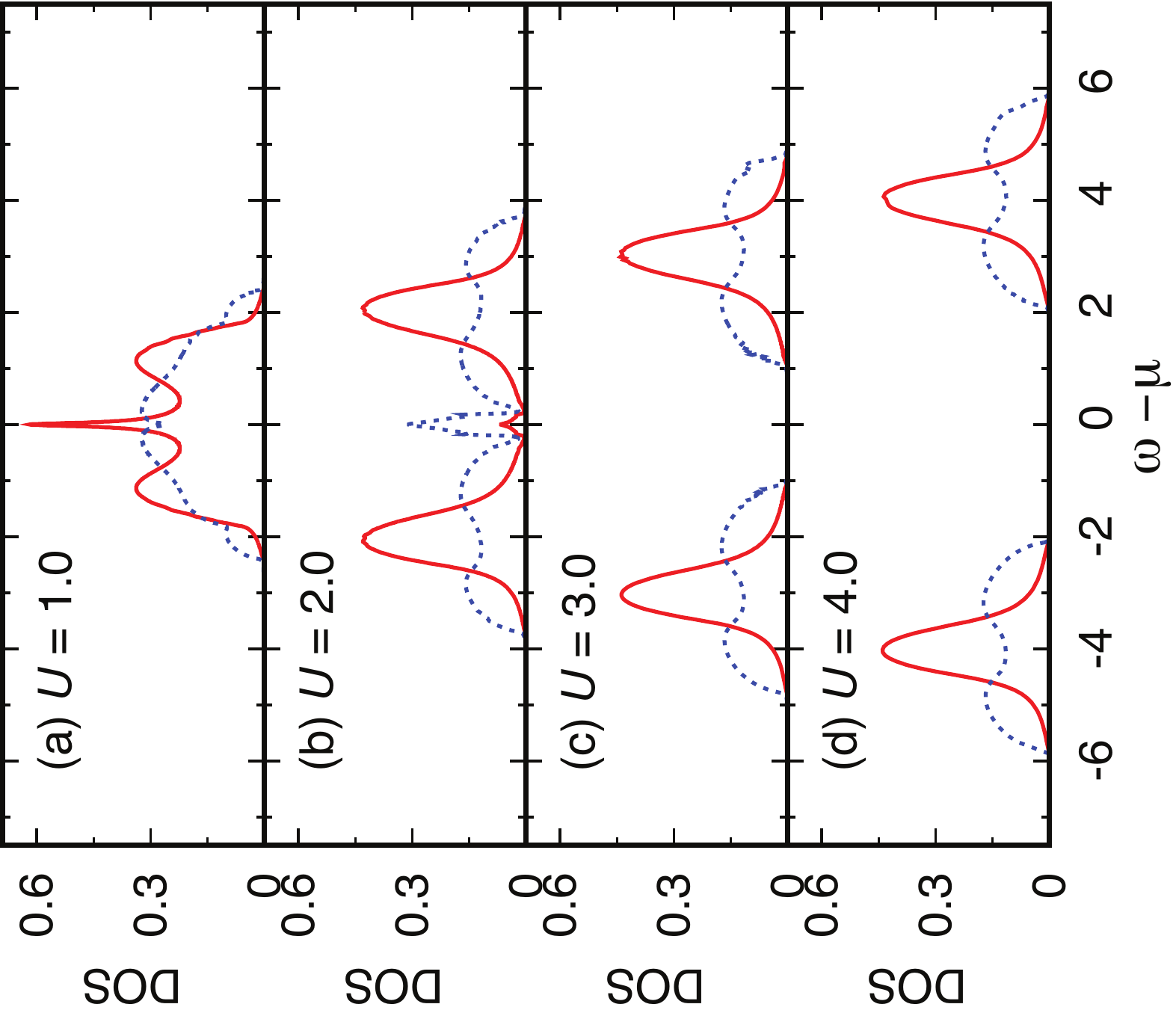}%
\end{center}
\caption{(Color online) Quasiparticle densities of states computed for the two-orbital system at halffilling for the case where the effect of hopping between orbitals in included. The results have been calculated with $D_2=2D_1=2$ and  $J=0$, and at $T=0.01$. The red solid line is for the narrow orbital, and the blue dashed line for the wide orbital.
 \label{figure6}}
\end{figure}

Next we study the case where we have  the halffilled two-orbital system with different bandwidths, but  identical intra-orbital and inter-orbital Coulomb interaction strengths, i.e., $J=0$. The obtained quasiparticle DOS is given in Figure~\ref{figure6}. The DOS reveals that the MIT for the two orbitals occurs nearly simultaneously. Compared with Figure~\ref{figure2}, the OSMT disappeared. This difference is due to the inclusion of the inter-site inter-orbital hopping effect. The critical value of $U$ for the two orbitals is here $U_{c}\approx 2.4$.

\begin{figure}[t!]
\begin{center}
\includegraphics[angle=-90,width=7.5cm]{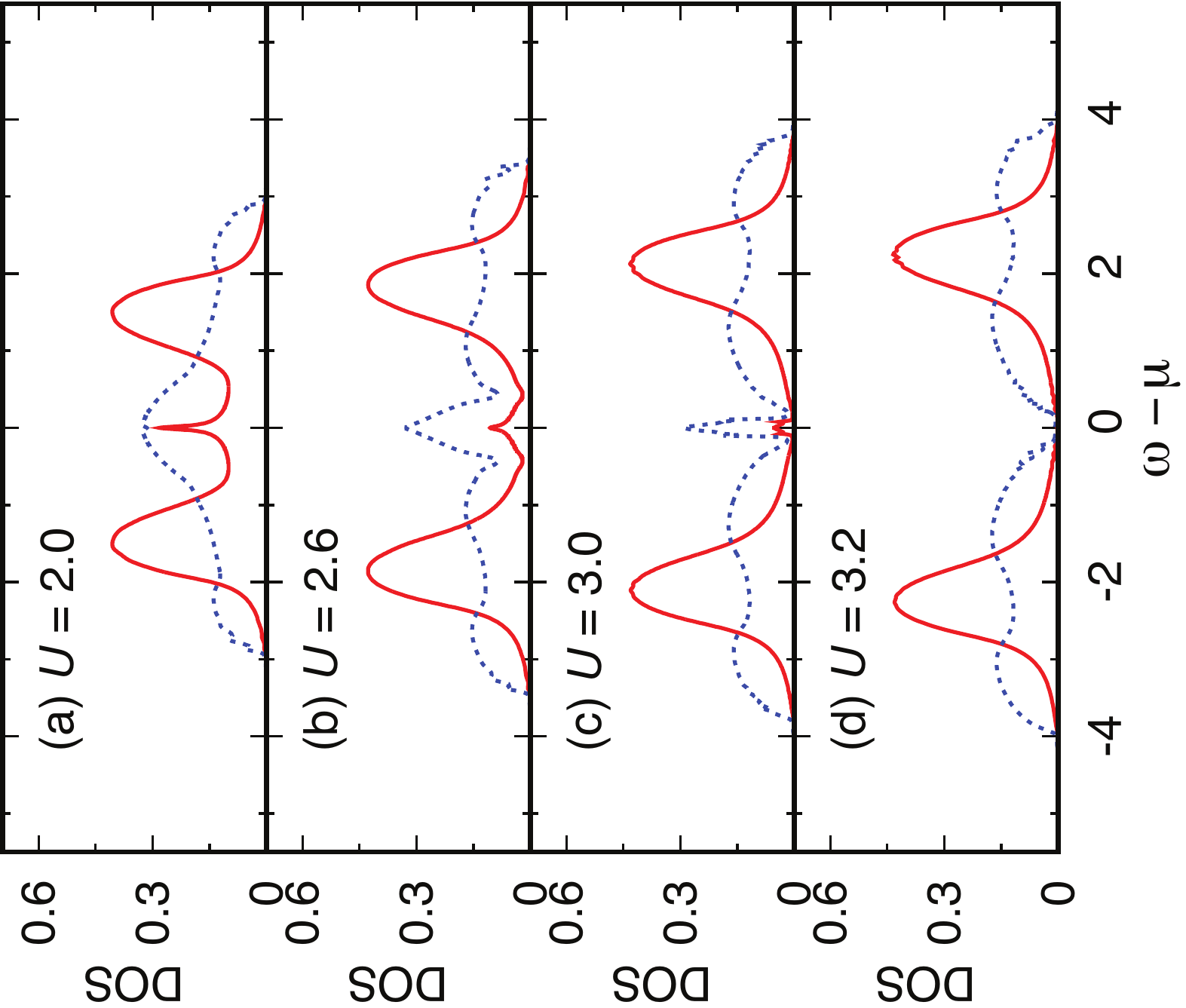}%
\end{center}
\caption{(Color online) Quasiparticle densities of states for the two-orbital system on the Bethe lattice at halffilling in case of including the effect of electron hopping between the two orbitals. The DOS have been obtained with $D_2=2D_1=2$, and $J=U/4$ and at $T=0.01$. The two orbitals are denoted by the dashed and full curves, as in Figure\ \ref{figure6}.
 \label{figure7}}
\end{figure}

When $J=U/4$ the computed DOS are shown in Figure~\ref{figure7}. We can see that the quasiparticle peak is nearly pinned at the chemical potential for the wide orbital. Near $U\approx 3.0$ the quasiparticle peak for the narrow band has become very small,
but it still exists. Compared with Figure~\ref{figure3}, the OSMT also disappeared under our present assumptions and on the Bethe lattice, which can not, however, exclude the OSMT in all cases, e.g., if the neighbour sites are different (i.e., the material is a compound) or inter-site multi-orbital hoppings are not like what we have assumed in a real system (physical lattices) or for magnetic cases when outer fields exist. But we believe that with the introduce of inter-site multi-orbital hopping effect, the OSMT will be suppressed in some sense.
This phenomena will be investigated in a further study with improved techniques beyond the here-made mean field approximation in the impurity solver and on other lattices.

\begin{figure}[tbh]
\begin{center}
\includegraphics[angle=-90,width=8cm]{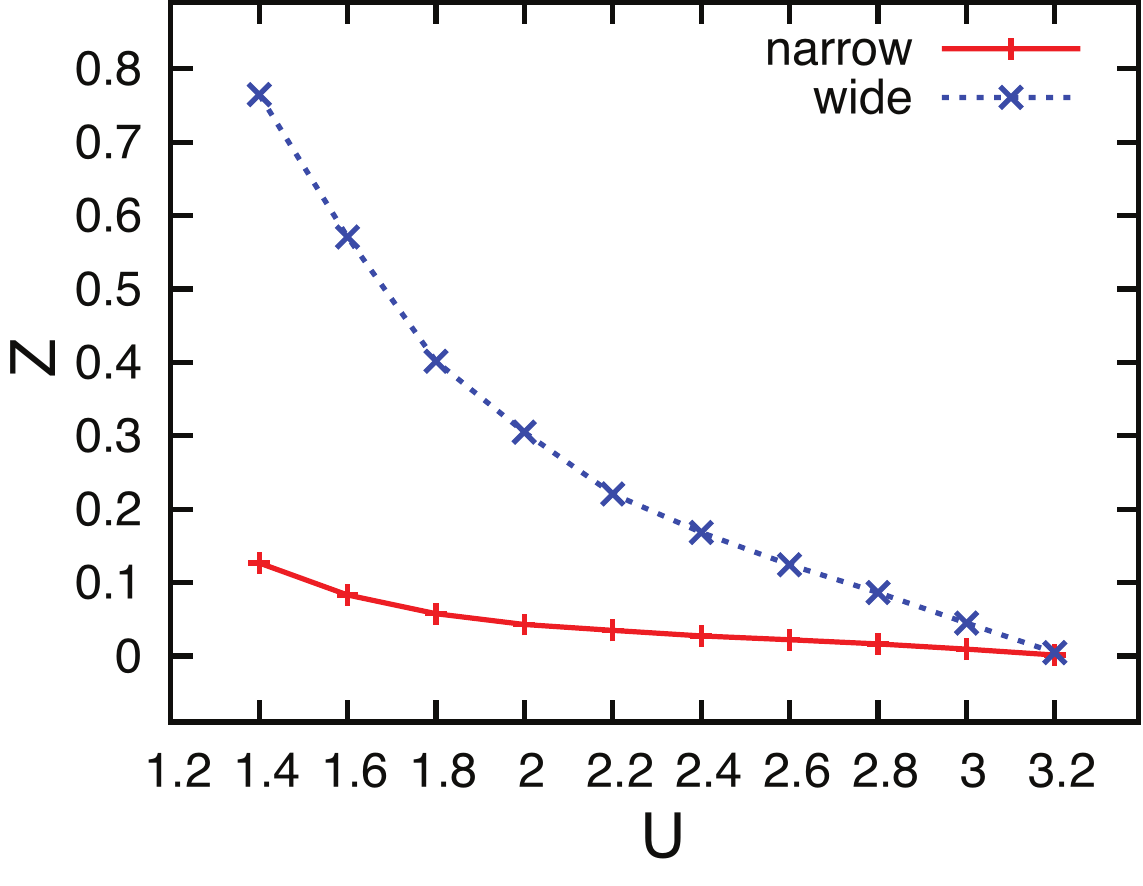}%
\end{center}
\caption{(Color online) Computed quasiparticle weight as a function of $U$ for the two-orbital system at halffilling, obtained with the parameters $D_2=2D_1=2$, $T=0.01$ and $J=U/4$.
\label{figure8}}
\end{figure}
Next we present the quasiparticle weight versus $U$ for the two-orbital system with different bandwidths at halffilling in Figure~\ref{figure8}.
We can clearly see in Figure\ \ref{figure8} that the two DOS curves of the wide and narrow orbitals approach zero at the same point, $U_c = 3.2$, which means that the metal-insulator transition for the two orbitals occurs practically simultaneously.

\begin{figure}[tbh]
\begin{center}
\includegraphics[angle=-90,width=8.5cm]{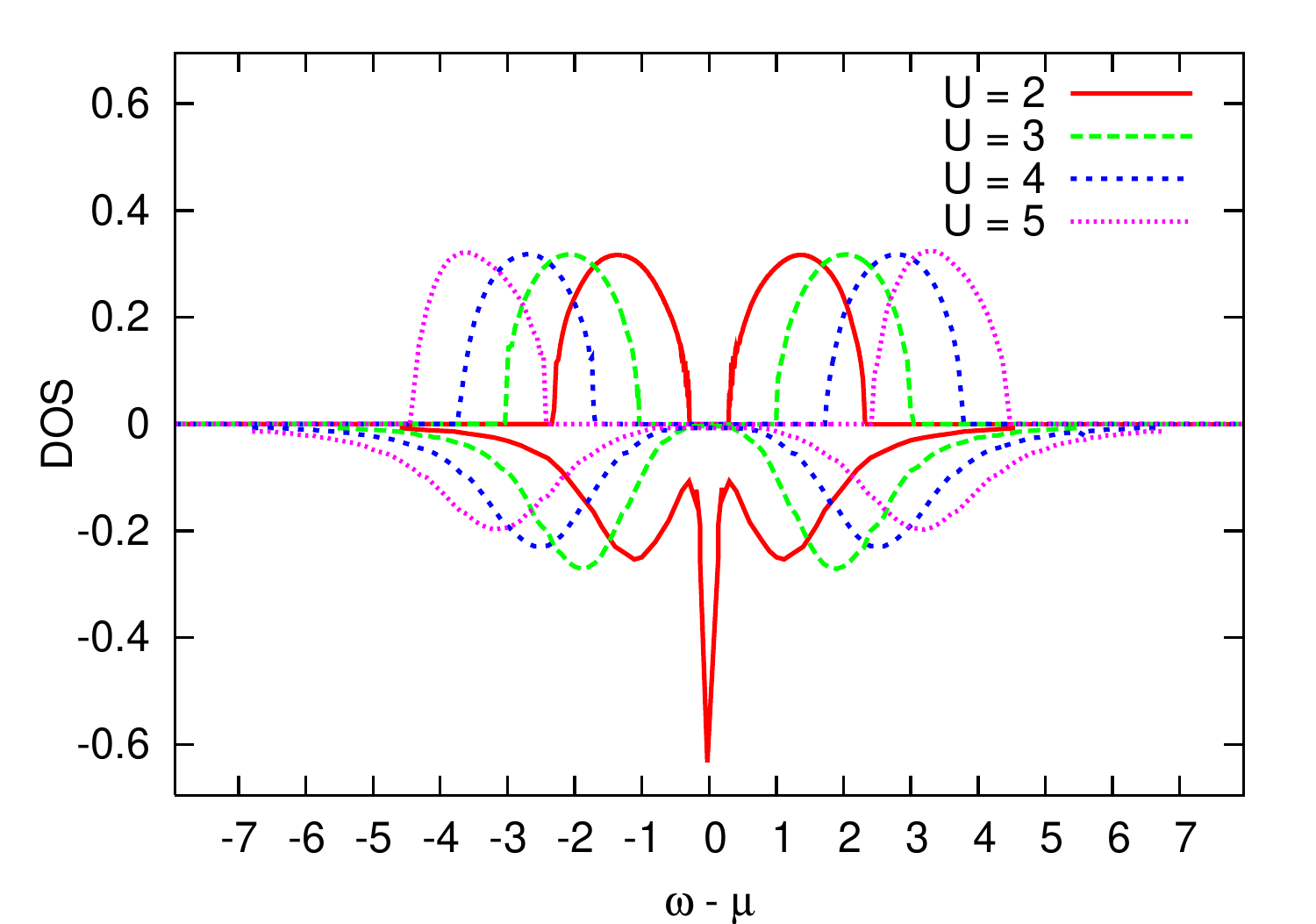}%
\end{center}
\caption{(Color online) Comparison of quasiparticle densities of states obtained by the multi-orbital EOM method with the DOS obtained by the NRG method. The system treated is a two-orbital system at halffilling on the Bethe lattice, with inclusion of the effect of electron hopping between orbitals with bandwidth $D_2=D_1=1$ and at temperature $T=0.01$, and $J=U/4$. The DOS shown on the negative ordinate are those for the NRG method, those on the positive ordinate for our MO-EOM method. The NRG data are taken from
\cite{PruschkeNRG}.
\label{figure9}}
\end{figure}
Next we present some comparisons of results computed with the MO-EOM method to results achieved with other numerical methods to see how well our MO-EOM impurity solver performs. One comparison is visualized in Figure~\ref{figure9}, where the DOS obtained with our MO-EOM method are compared with those achieved with the NRG method \cite{PruschkeNRG} for the case that the two orbitals have identical band widths. One can note that both the methods agree well with the positions of the Hubbard bands. They do show a difference in the critical value of $U$ for the Mott metal-insulator transition, which can however be partly explained by the large Lorentzian broadening (which causes a long tail of the peak) shown in the NRG data. In our genetic algorithm method this Lorentzian broadening can even be set as zero.  Considering this broadening effect, our results are close to those obtained with the NRG method, but may have a slightly larger critical value of $U$.

\begin{figure}[b!]
\begin{center}
\includegraphics[angle=-90,width=8.5cm]{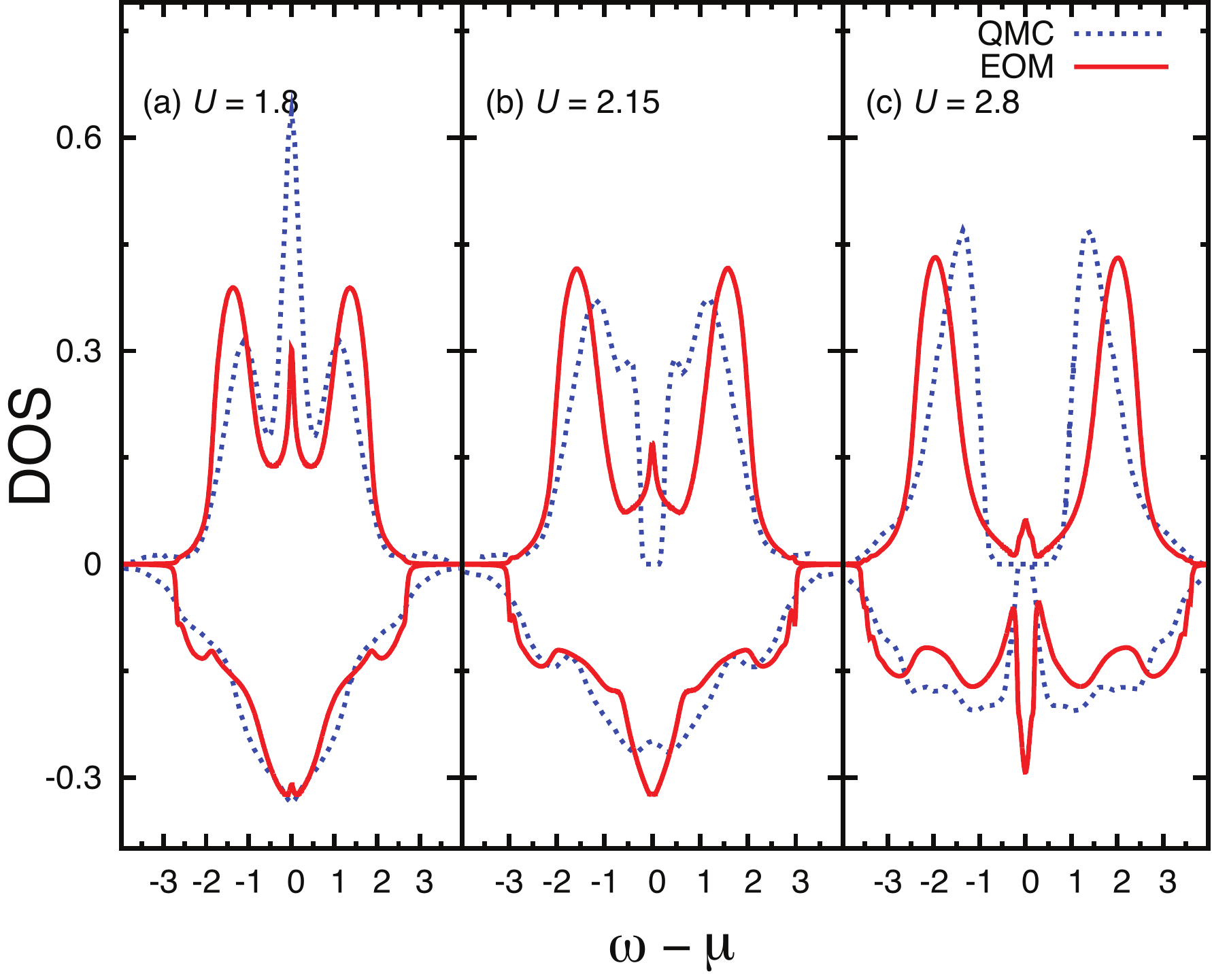}%
\end{center}
\caption{(Color online) Comparison of quasiparticle densities of states obtained by the multi-orbital EOM method with DOS obtained by the QMC method. The system treated is a two-orbital system at halffilling on the Bethe lattice, with inclusion of the effect of electron hopping between orbitals with bandwidth $D_2=2D_1=2$ and at temperature $T=1/40$, and $J=U/4$. The DOS shown on the negative ordinate are those for the wide orbital, those on the positive ordinate for the narrow orbital. The QMC data are taken from
\cite{OSMT1} and all parameters are identical to those of \cite{OSMT1}.
\label{figure10}}
\end{figure}
Next we present a comparison of results computed with the MO-EOM method to results achieved with the QMC method \cite{OSMT1}. This comparison is visualized in Figure~\ref{figure10}, though the Hamiltonian we studied here involved the inter-orbital hybridizations, a little different to that used in \cite{OSMT1}.
For visibility purpose, we have plotted the quasiparticle  DOS of the wide orbital on the
negative $y$ axis and that of the narrow orbital on the positive $y$ axis. One can recognize a good agreement between the two DMFT solvers, particularly for the lower $U$ values. For the higher $U$ value there is some difference, which is related to the fact that the critical value $U_c$ in our calculations is larger than in the QMC results. This may however be due to the different Hamiltonians used in the simulations. We used a less exact approximation (mean field approximation to inter-orbital Coulomb interactions) but a more realistic Hamiltonian, while the QMC method used an exact numerical method but a simplified Hamiltonian. Therefore here we can not conclude which one is more precise. In our opinion, it is reasonable to obtain a larger critical value of $U$ in our MO-EOM method, because we have taken into account the hopping between different orbitals of different sites. Moreover, in our treatment we have assumed that the probability of electrons hopping from the narrow to the wide orbital is $\frac{t_1t_2}{(t_1+t_2)}$ which is much larger than the probability $\frac{t^2_1}{(t_1+t_2)}$ of hopping from one narrow to another narrow orbital. If more electrons would be allowed to hop into an orbital with identical orbital index, the critical value of $U$ will become smaller. We should also note that the mean field approximation made for the inter-orbital Coulomb interactions might have narrowed the gap between lower and upper Hubbard band, i.e., this would lead to an increase of the critical value of $U$.
Using the present mean field approximation to achieve comparable results as the QMC method, the future extension beyond the mean field approximation is promising.

\begin{figure}[b!]
\begin{center}
\includegraphics[angle=-90,width=8.5cm]{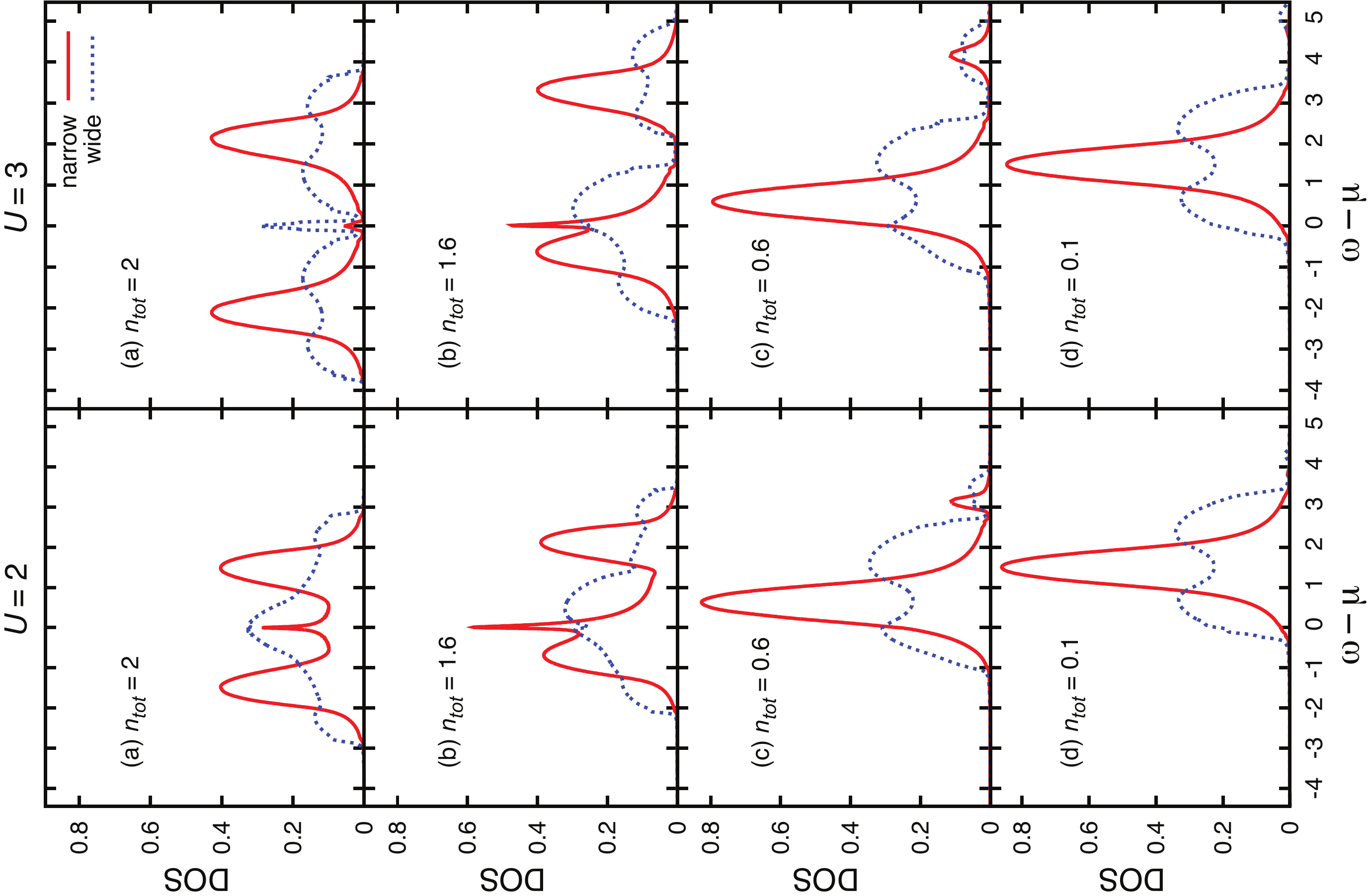}%
\end{center}
\caption{(Color online) Quasiparticle densities of states computed for the two-orbital system on the Bethe lattice for different total fillings. The parameters used are $D_2=2D_1=2$, $T=0.01$, and $J=U/4$. The red solid line is for the narrow orbital, and the blue dashed line for the wide orbital.
\label{figure11}}
\end{figure}

As a next step we investigate the influence of the filling level. To this end we have varied the total occupation number in the simulations. We present the computed DOS of the two-orbital systems at different occupation numbers in Figure~\ref{figure11}, where we have taken the parameters that the two orbital levels are identical, $i.e.,~E_1=E_2$.
Vertically, the figure shows the changes of the DOS along with the total occupancy. When the filling of the orbitals increases, the effective Coulomb interaction is seen to become larger.
Correspondingly, the DOS can be seen to change from a one or two-peak structure to a three peak structure typical of strongly correlated systems, where the quasiparticle peak appears at the Fermi level. In this procedure, an orbital selective phenomena is observed, i.e., at the beginning the occupation in the narrow orbital is less than that in the wide orbital, then changes to equal and then larger than the occupation in the wide orbital, finally equal again at the half-filling, $n_{tot}=2$. If the intra-orbital Coulomb interaction strengths or orbital levels or both are different for the two orbitals, the change of occupation and effective Coulomb interaction in the two orbitals will be more complicated. In addition we compare in Figure\ \ref{figure11} the DOS calculated with different Coulomb interaction strengths, $U=2$ and $U=3$. It can be recognized that, when $U$ increases, the two Hubbard bands move farther away from each other, as anticipated, and a gap appears in the unoccupied spectrum at, e.g., $\omega-\mu\approx2$ for $U=3$ and $n_{\it tot}= 1.6$. At $n_{\it tot} = 2$, the quasiparticle peak is also reducing along with the increment of $U$. When $U$ is large enough, the quasiparticle peak will eventually disappear and a gap appears at the chemical potential, as shown in panel (d) of Figure~\ref{figure7}, and the system will turn to an insulating state.

\begin{figure}[h!]
\begin{center}
\includegraphics[angle=-90,width=8cm]{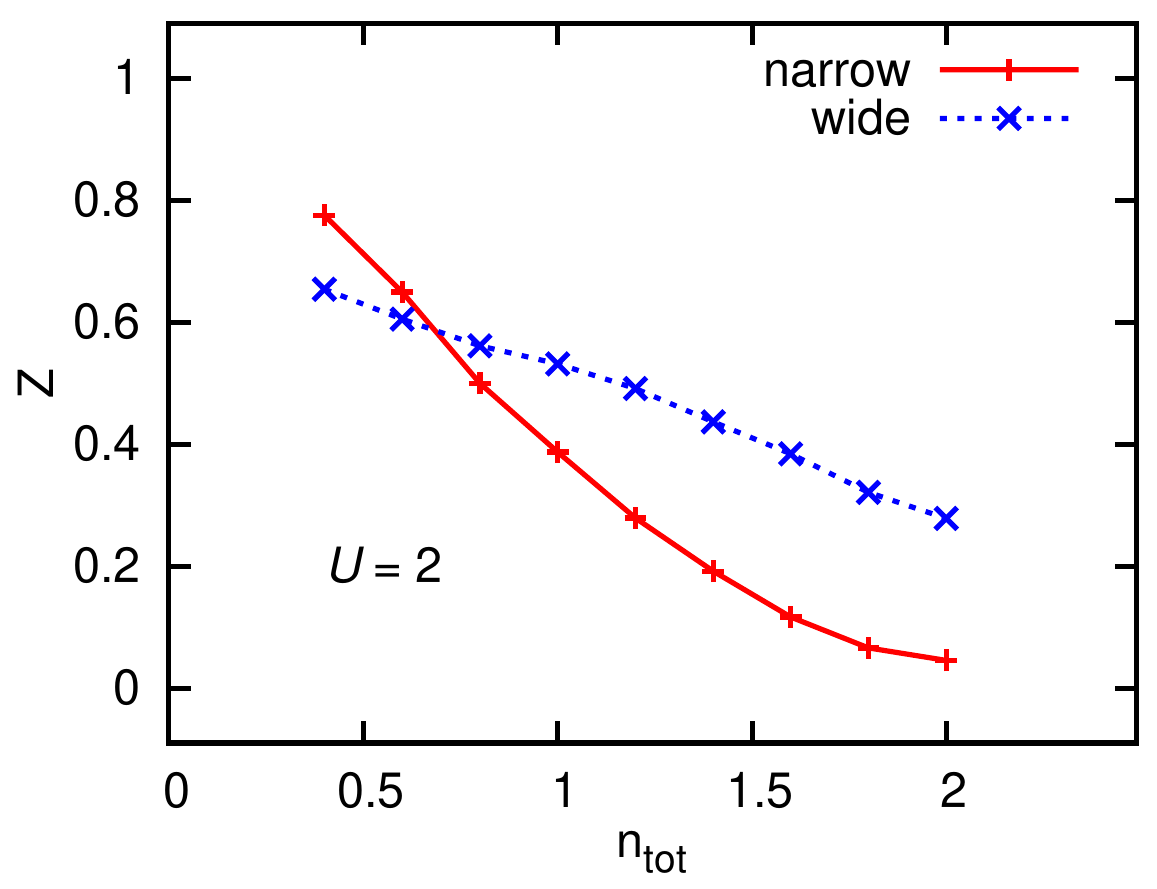}%
\end{center}
\caption{(Color online) Quasiparticle weights of the narrow and wide orbitals as a function of the total filling for the two-orbital system on the Bethe lattice, computed with the parameters $D_2=2D_1=2$, $T=0.01$, and $J=U/4$.
\label{figure12}}
\end{figure}
 Figure~\ref{figure12} shows the quasiparticle weight versus the total occupation number of the two orbitals. Due to the fact that $J=U/4$ and that the occupation numbers in the two orbitals are different, the effective Coulomb interaction for the electrons in the narrow orbital is smaller than that for electrons in the wide orbital when $n_{tot}$ is small, this causes the narrow orbital to have a larger quasiparticle weight than the wide orbital. Yet, when the occupation of the narrow orbital increase, its weight becomes reduced, leading to a cross-over point of the quasiparticle weights of the two orbitals.

\section{Summary\label{sect4}}
In this paper we have proposed and tested a fast multi-orbital impurity solver for the DMFT based on the equations of motion method.  In the construction we have introduced  a mean field approximation for inter-orbital Coulomb interactions appearing with the two and three particle Green's function expression.  The proposed impurity solver includes inter-orbital hybridizations and it preserves the particle-hole symmetry for halffilled states, and it automatically satisfies the sum rule that integrating of the density of states
over all $\omega$ space for each spin and orbital index equals identity.
The influence of the inter-orbital hopping has been investigated numerically with the MO-EOM method.
The results obtained with our impurity solver show that the inclusion of inter-site inter-orbital hopping does make a difference and indicate that the inter-site inter-orbital hopping effect is quite important for multi-orbital systems. It is observed that the inclusion of the inter-site inter-orbital hopping effect will strengthen the connections between different sites (and hence eliminate the difference between the orbitals if the neighbour sites are all identical to this impurity), and has caused the disappearance of the OSMT under our present assumptions and on the Bethe lattice. This will be tested further in a forthcoming work with treatments beyond the mean field approximation and on different lattices.
Although our approximations are simplifying in some respects, we do believe that, if one wants to study the difference between orbitals having different bandwidths, e.g., the OSMT, the inter-site inter-orbital hopping can not be neglected. 

Moreover, we note that the decoupling scheme only acts on the on-site inter-orbital fluctuation terms of the impurity, and does not influence the spatial fluctuations of the lattice model.
The spatial fluctuations should be considered in the DMFT self-consistency conditions or a future EOM cellular DMFT.

The precision of the EOM method depends on the decoupling only.  A drawback of this impurity solver is that the mean field approximation to the on-site inter-orbital Coulomb interactions may give rise to the loss of some interesting information when the on-site inter-orbital fluctuations are strong. Hence, if one wants to study the inter-orbital fluctuations and even phenomena related to higher-order terms (e.g., spin-flip term and pair hopping term), a higher-order decoupling scheme will be needed. Nonetheless, the present EOM method can already be applied to most systems.

The main advantages of the MO-EOM method are, first, that it works directly on real frequencies and for arbitrary temperatures,
second, it is a fast impurity solver for the multi-orbital case. Finally, it can work for an arbitrary number of orbitals.
Consequently, if one could determine all the parameters between the orbitals properly, even all the orbitals of one atom could be treated within MO-EOM method.
Therefore the developed MO-EOM method is a competitive solver as compared, e.g., to the QMC.
The multi-orbital EOM solver can be applied to multi-orbital system containing several orbital levels, with varying band widths, inter-orbital and intra-orbital Coulomb interaction strengths.
Also systems with a crystal field  having different orbital levels can be treated with this MO-EOM method.
Besides the model calculation, using this impurity solver and the formula that we have given for orbital levels and effective Coulomb strength, one can obtain the occupations and then use these to correct the charge, orbital levels and remove the double counting, 
so that it can be implemented in DFT+DMFT calculations for real correlated materials.

In the peer review of this paper, our attention was drawn to the paper by  P.\ Hansmann {\it et al.}, who have mentioned the inter-site inter-orbital hoppings in their LDA+DMFT work. But these have however not appeared in model calculations to discuss the influence of these hoppings on the orbital selective Mott transition.
\ack
We would like to thank Y.-Z.\ Zhang and H.\ O.\ Jeschke for discussions in an early stage of this work. We acknowledge financial support from the Swedish Research Council (VR) and from SKB. Part of 
the calculations have been supported through the Swedish National Infrastructure for Computing (SNIC) and have preformed at the Swedish national computer centers NSC, UPPMAX, and HPC2N.


\section*{References}
\begin{thebibliography}{99}

\bibitem{DMFT1} Metzner W and Vollhardt D 1989 {\it Phys. Rev. Lett.} {\bf 62}, 324

\bibitem{DMFT2} Georges A and Kotliar G 1992 {\it Phys. Rev. B} {\bf 45}, 6479

 \bibitem{DMFTRMP96} Georges A, Kotliar G, Krauth W and Rozenberg M J 1996 {\it Rev. Mod. Phys.} {\bf 68} 13

 \bibitem{DMFTRMP06} Kotliar G, Savrasov S Y, Haule K, Oudovenko V S and Parcollet O 2006 {\it Rev. Mod. Phys.} {\bf 78} 865

 \bibitem{HF-QMC} Hirsch  J E and Fye R M 1986 {\it Phys. Rev. Lett.} {\bf 56} 2521

 \bibitem{ED} Caffarel M and Krauth W 1994 {\it Phys. Rev. Lett.} {\bf 72} 1545

 \bibitem{ED2} Si Q, Rozenberg M J, Kotliar G and Ruckenstein A E 1994 {\it Phys. Rev. Lett.} {\bf 72} 2761

 \bibitem{NRG} Bulla R, Hewson A C and Pruschke Th 1998 {\it J. Phys.: Condens. Matter} {\bf 10} 8365

 \bibitem{NRG2} Bulla R 1999 {\it Phys. Rev. Lett.} {\bf 83} 136

 \bibitem{DMRG} Garcia D J, Hallberg K and Rozenberg M J 2004 {\it Phys. Rev. Lett.} {\bf 93} 246403

 \bibitem{CTQMC} Rubtsov A N, Savkin V V and Lichtenstein A I 2005 {\it Phys. Rev. B} {\bf 72} 035122

 \bibitem{CTQMC2} Werner P, Comanac A, Medici L De', Troyer M and Millis A J 2006 {\it Phys. Rev. Lett.} {\bf 97} 076405

 \bibitem{Hubbard63} Hubbard J 1963 {\it Proc. Roy. Soc. A} {\bf 276} 238

 \bibitem{Hubbard64a} Hubbard J 1964 {\it Proc. Roy. Soc. A} {\bf 277} 237

 \bibitem{Hubbard64b} Hubbard J 1964 {\it Proc. Roy. Soc. A} {\bf 281} 401

 \bibitem{Lacroix1} Laxroix C 1981 {\it J. Phys. F: Met. Phys.} {\bf 11} 2389

 \bibitem{Lacroix2} Lacroix C 1982 {\it J. Appl. Phys.} {\bf 53} 2131

 \bibitem{Czycholl85} Czycholl G 1985 {\it Phys. Rev. B} {\bf 31} 2867

 \bibitem{Petru93} Petru J 1993 {\it Z. Phys. B} {\bf 91} 351

 \bibitem{GrosEOM} Gros G 1994  {\it Phys. Rev. B} {\bf 50} 7295

 \bibitem{Kang95} Kang K and Min B I 1995 {\it Phys. Rev. B} {\bf 52} 10689

 \bibitem{Luo99} Luo H G, Ying Z J and Wang S J 1999 {\it Phys. Rev. B} {\bf 59} 9710

 \bibitem{ZhuEOM1} Zhu J -X, Albers R C and Wills J M 2006 {\it Mod. Phys. Lett. B} {\bf 20} 1629

 \bibitem{ZhuEOM2} Qi Y, Zhu J -X and Ting  C S 2009 {\it Phys. Rev. B} {\bf 79} 205110

 \bibitem{OPM1} Onoda S and Imada M 2001 {\it J. Phys. Soc. Jpn.} {\bf 70} 632

 \bibitem{OPM2} Onoda S and Imada M 2001 {\it J. Phys. Soc. Jpn.} {\bf 70} 3398

 \bibitem{JK05} Jeschke H O and Kotliar G 2005 {\it Phys. Rev. B} {\bf 71} 085103

 \bibitem{FZJ09} Feng Q, Zhang Y -Z and Jeschke H O 2009 {\it Phys. Rev. B} {\bf 79} 235112

 \bibitem{Dai} Zhuang J N, Wang L, Fang Z and Dai X 2009 {\it Phys. Rev. B} {\bf 79} 165114

 \bibitem{phdthesis-feng} Feng Q 2009 {\it Ph.D. thesis, University Frankfurt}

 \bibitem{OSMT1} Knecht C, Bl\"umer N and Dongen P G J van 2005 {\it Phys. Rev. B} {\bf 72} 081103R

 \bibitem{Zubarev} Zubarev D N 1960 {\it Sov. Phys. Usp.} {\bf 3} 320

 \bibitem{CFCT09} Capone M, Fabrizio M, Castellani C and Tosatti E 2009 {\it Rev. Mod. Phys.} {\bf 81} 943

 \bibitem{Yin11} Yin Y P, Haule K and Kotliar G 2011 {\it Nature Physics} {\bf 7}, 294 

 \bibitem {PruschkeNRG} Pruschke Th and Peters R 2007 {\it J. Magn. Magn. Mater.} {\bf 310}, 935

 \bibitem{Held} Hansmann P, Toschi A, Yang X, Andersen O K, and Held K 2010 {\it Phys. Rev. B} {\bf 82}, 235123
\end{thebibliography}
\end{document}